\documentclass[journal]{IEEEtran}
\usepackage{ifpdf}

\usepackage{cite}

\ifCLASSINFOpdf
   \usepackage[pdftex]{graphicx}
   \graphicspath{{../Figures/}}
   \DeclareGraphicsExtensions{.pdf}
\else
\fi

\usepackage{amsfonts}
\usepackage{amsmath}
\usepackage{algorithm}
\usepackage{algorithmic}

\usepackage{array}

\ifCLASSOPTIONcompsoc
	\usepackage[caption=false,font=normalsize,labelfont=sf,textfont=sf]{subfig}
\else
	\usepackage[caption=false,font=footnotesize]{subfig}
\fi

\usepackage{url}

\makeatletter
\let\NAT@parse\undefined
\makeatother
\usepackage{hyperref}  
\usepackage{color}

\begin{document}

\title{RIS-Aided Wireless Communications: \\ Prototyping, Adaptive Beamforming, and Indoor/Outdoor Field Trials}

\author{Xilong Pei, Haifan Yin, Li Tan, Lin Cao, Zhanpeng Li, Kai Wang, Kun Zhang, and Emil Bj{\"o}rnson
\thanks{X. Pei, H. Yin, L. Tan, L. Cao, Z. Li, K. Wang, K. Zhang are with the School of Electronic Information and Communications, Huazhong University of Science and Technology, Wuhan, China. E-mail: \{pei, yin, ltan, caolin, zhanpli, kaiw, zkun\}@hust.edu.cn.}
\thanks{E. Bj\"ornson is with the KTH Royal Institute of Technology, Sweden, and Link\"oping University, Sweden. E-mail: emilbjo@kth.se.}
\thanks{The corresponding author is Haifan Yin.}
\thanks{This work was supported by the National Natural Science Foundation of China under Grant 62071191. E. Bj\"ornson was supported by the FFL18-0277 grant from the Swedish Foundation for Strategic Research.}}


\maketitle

\begin{abstract}

The prospects of using a Reconfigurable Intelligent Surface (RIS) to aid wireless communication systems have recently received much attention from academia and industry. Most papers make theoretical studies based on elementary models, while the prototyping of RIS-aided wireless communication and real-world field trials are scarce. In this paper, we describe a new RIS prototype consisting of 1100 controllable elements working at 5.8 GHz band. We propose an efficient algorithm for configuring the RIS over the air by exploiting the geometrical array properties and a practical receiver-RIS feedback link. In our indoor test, where the transmitter and receiver are separated by a 30 cm thick concrete wall, our RIS prototype provides a 26 dB power gain compared to the baseline case where the RIS is replaced by a copper plate. A 27 dB power gain was observed in the short-distance outdoor measurement. We also carried out long-distance measurements and successfully transmitted a 32 Mbps data stream over 500 m. A 1080p video was live-streamed  and it only played smoothly when the RIS was utilized. The power consumption of the RIS is around 1 W. Our paper is vivid proof that the RIS is a very promising technology for future wireless communications.

\end{abstract}


\IEEEpeerreviewmaketitle

\section{Introduction}

Massive multiple-input multiple-output (MIMO) technology \cite{marzetta:10a, 6736761,massivemimobook} is at the core of the 5th generation (5G) cellular communication technology.
The base stations make use of arrays of 64 or more antenna-integrated radios \cite{AASbook} to enable precise beamforming towards any location in the cell and to enable spatial multiplexing of many user terminals.
The technology can focus energy in space but it cannot overcome major weaknesses imposed by the propagation environment, such as shadow fading.
The losses by shadowing particularly evident when the carrier frequency increases, e.g., from the 3.5 GHz 5G band to millimeter wave (mmWave) bands, and even towards the (sub-)terahertz bands.
The traditional approach to overcome such propagation limitations is to utilize relays in between the base station and intended receivers to fill coverage holes \cite{Dohler2010a}. This equipment could be anything from passive repeaters (e.g., a carefully rotated copper plate that reflects signals in a predetermined direction) to relays with baseband processing (e.g., using a decode-and-forward protocol where an amplified signal is retransmitted after noise removal).

A new type of relaying technology has appeared inspired by recent advances in electromagnetic metamaterials \cite{Liaskos2018a,Renzo2020b,Bjornson2020,9326394}. It is known as a Reconfigurable Intelligent Surface (RIS) \cite{Huang2018a} or Intelligent Reflecting Surface \cite{Wu2019}.
In a nutshell, an RIS is a two-dimensional surface that can be electronically tuned to interact with electromagnetic waves as if it had another physical shape; for example, if one wants the transmitted signal to be reflected towards a certain location, the RIS can synthesize a metal plate that is rotated and bent to focus the incident waves on that location \cite{RIS_SPMAG}.
Practically speaking, an RIS can be built using artificial electromagnetic metamaterial, which consists of periodic arrangements of specifically designed subwavelength-sized structural elements \cite{cui2009ruopeng}.
Such metamaterials have unique electromagnetic properties that do not exist in nature \cite{cui2017information}, such as, negative refraction \cite{shelby2001experimental}, perfect absorption, and anomalous reflection/scattering \cite{liang2019large}.
In order to realize the aforementioned relaying feature, the RIS must contain a large number of elements (to obtain appreciable gains) that have controllable properties \cite{Tsilipakos2020a}. By varying the reflection coefficient (e.g., phase shift) of the elements, one can in theory control towards which location an incident wave is beamformed. The number of discrete states that an element can have is often measured in bits, because an element with $2^N$ states can be controlled using $N$ bits \cite{Wu2020b}.

We will put the concept of RIS to practice by designing an RIS-enhanced wireless communication prototype and using it for indoor/outdoor field trials.
A previous prototype was described in \cite{Tang2019Wireless, Tang2019Programmable, 9048622}. It was designed to modulate the impinging signal, which is very different from our work, where we passively reflect/relay information-carrying wireless signals towards the receiver.
In \cite{Dai2020}, an RIS prototype with 256 2-bit elements was presented. It achieved a 21.7 dBi antenna gain for a fixed configuration and no adaptive beamforming was reported.
The power consumption was around 153 W.
\cite{arun2020rfocus} proposed RFocus, which is a system that moves beamforming functions from the radio transmitter to the environment. Their measurements showed an improvement in median signal strength by 9.5 times and doubled the median channel capacity. The beamforming functionality relied on signal strength measurements, which is the Received Signal Strength Indication (RSSI).
In \cite{ozdemir202028}, Metawave Corporation demonstrated a 5G coverage extension use case with a 28 GHz passive reflector.

In this paper, we describe a prototype of RIS-aided wireless communication system that can serve mobile users through feedback-based real-time beamforming at the RIS.
Our system operates in the 5.8 GHz band, which is the unlicensed band used by wireless local area network (WLAN).
The host computer of the transmitter encodes a video stream to be transmitted.
The transmitter carries out baseband signal processing and radio-frequency (RF) signal transmission through a horn antenna.
The RIS controls the reflection coefficients by adjusting the impedance of the patches and thereby beamforming the impinging signal towards the receiver.
The receiver processes the baseband signal and demodulates the  video stream.
In the meantime, the receiver calculates the strength of the received signal and feeds back this information to the RIS.
The RIS updates the reflection coefficients according to a proposed self-adaptive beamforming algorithm.
The power consumption of our RIS is only about 1 W.

The main contributions of our work are as follows:

\begin{itemize}
	\item We have developed a prototype for RIS-aided communication and evaluated its basic properties in the lab.
	
	\item We verify the uplink and downlink channel reciprocity of the RIS communication channel.
	We also demonstrate the reflection phase response for a given control voltage varies with the incident angle of the electromagnetic wave. This non-ideal feature must be considered in practical designs.
	
	\item We have proposed and tested a novel feedback-based adaptive reflection coefficient optimization algorithm, which enables smart reflection without having to modify the existing communication standards. 
	
	\item We conduct the world's first indoor test without a Line-of-Sight (LoS) path and demonstrate a 26 dB power gain,  compared to having a copper plate that lacks the capability of smart reflection.
	
	\item We conduct the world's first long-range outdoor test.
	In the 50\,m and 500\,m field trials, the RIS achieved up to 27 dB power gains. In the longest scenario, a 32 Mbps data rate was achieved with the help of the RIS.
	
\end{itemize}

The rest of this paper is organized as follows. Section \ref{sec:board} describes the details of our RIS design.
In Section \ref{sec:system}, we show the prototype of an RIS-based wireless communication system.
In Section \ref{sec:codebook}, we focus on the design of the reflection coefficients.
Then in Section \ref{sec:results}, we make several experiments to verify the system performance.
Finally, the conclusions are drawn in Section \ref{sec:conclusion}.

\section{Design of Reconfigurable Intelligent Surface}\label{sec:board}

In this section, we describe the details of our new large-scale RIS hardware prototype, which consists of 1100 elements.

\subsection{Reflective Element Design}

\begin{figure}[t!]
	\centering
	\subfloat[Perspective view]{\includegraphics[width=0.5\linewidth]{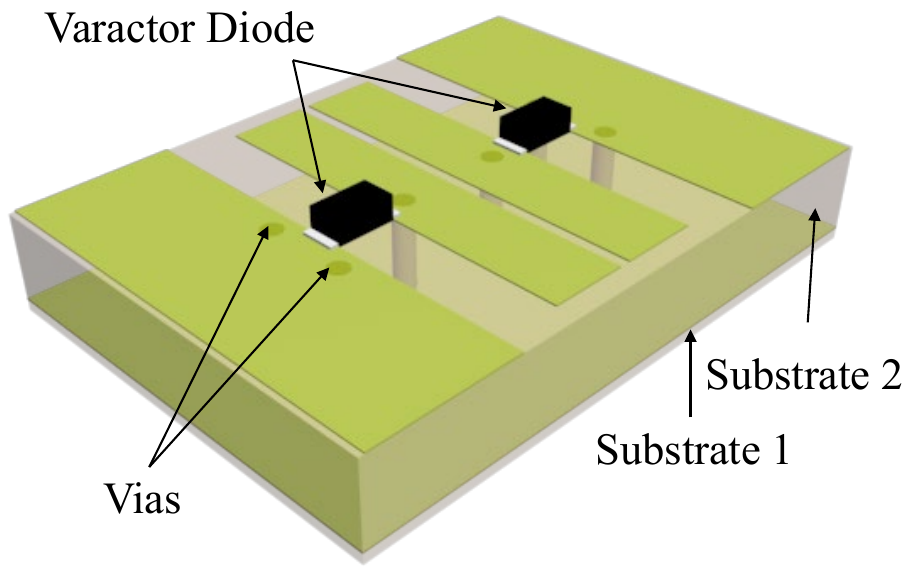}%
		\label{fig:PerspectiveView}}
	\hfil
	\subfloat[Top view]{\includegraphics[width=0.5\linewidth]{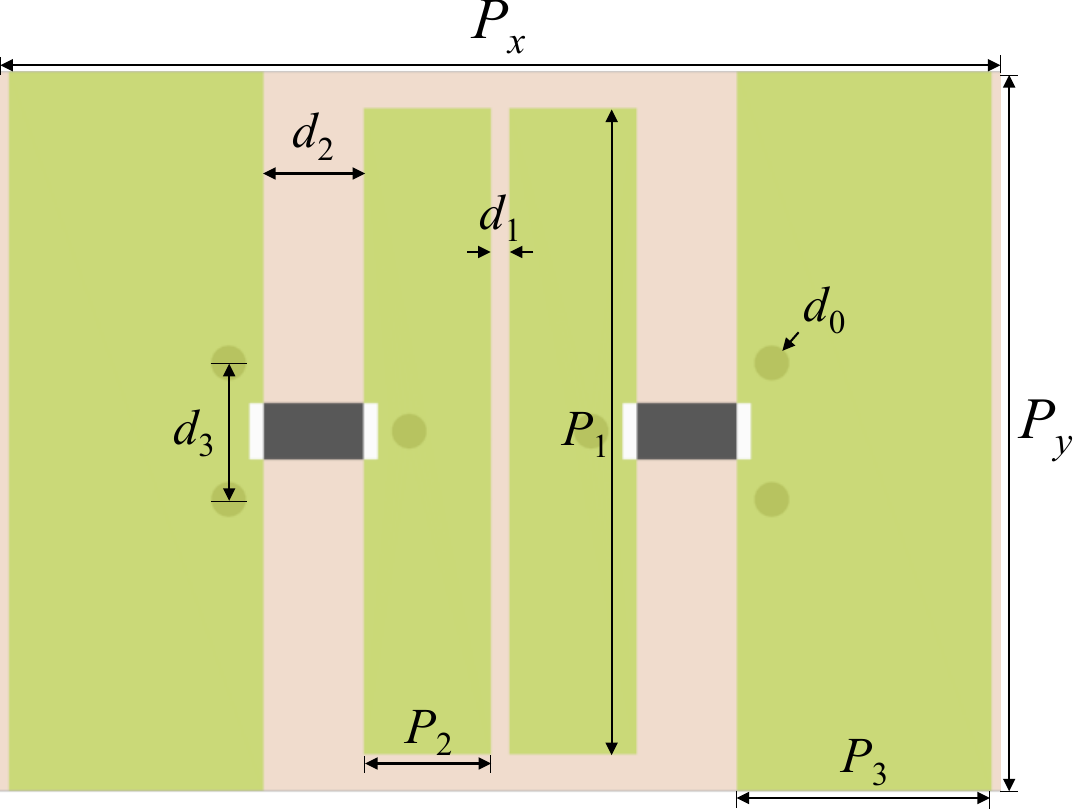}%
		\label{fig:TopView}}
	\hfil
	\subfloat[Side view]{\includegraphics[width=0.6\linewidth]{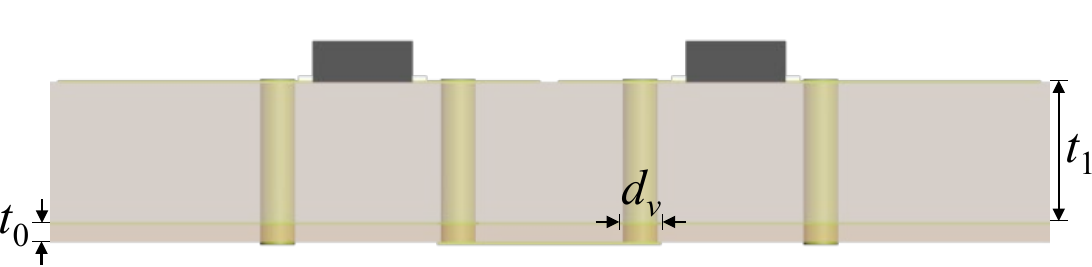}%
		\label{fig:SideView}}
	\caption{The topology of an RIS element. Each element contains two varactor diodes, and the parameters are marked in the figure.}
	\label{fig:Topology}
\end{figure}

\begin{table}[t!]
	\renewcommand{\arraystretch}{1.3}
	\caption{The Structural Parameters of an RIS Element}
	\label{tab:RISparameters}
	\centering
	\scriptsize
	\begin{tabular}{|c|c|}
		\hline\hline
		  \textbf{Parameter}   &                               \textbf{Value}                                \\ \hline
		        Period         &            $P_x=14.3 \, \mathrm{mm}, P_y = 10.27 \, \mathrm{mm}$            \\ \hline
		Rectangular patch size & $P_1=9.23 \, \mathrm{mm}, P_2=1.82 \, \mathrm{mm}, P_3=3.64 \, \mathrm{mm}$ \\ \hline
		   Patch slot width    &             $d_1=0.26 \, \mathrm{mm}, d_2=1.43 \, \mathrm{mm}$              \\ \hline
		 Via-holes clearance   &                         $ d_3=1.95 \, \mathrm{mm} $                         \\ \hline
		 Ground hole opening   &                         $ d_v=0.6 \, \mathrm{mm} $                          \\ \hline
		     Via diameter      &                         $ d_0=0.5 \, \mathrm{mm} $                          \\ \hline
		     Substrate 1       &  FR4($\varepsilon _r=4.4, \tan \delta =0.02, t_0 = 0.254 \, \mathrm{mm}$)  \\ \hline
		     Substrate 2       &   F4B ($\varepsilon _r=2.65, \tan \delta =0.005, t_1 = 2 \, \mathrm{mm}$)   \\ \hline\hline
	\end{tabular}
	\normalsize
\end{table}

We designed a metasurface tuned by varactor diodes for operation in the C-band.
\figurename{~\ref{fig:Topology}} shows a varactor diode tuned structure. 
The structure of the designed reflective element is inspired by \cite{dai2019wireless}, with an adaption to the desired frequency.
The structural parameters are listed in Table {\ref{tab:RISparameters}} and their meanings are shown in \figurename{~\ref{fig:Topology}}.
The reflective elements are densely packed without spacing. 
Each element consists of three layers.
The top layer contains two pairs of rectangular metallic patches, each of which is connected by a varactor diode with the junction capacitance controlled by the external bias voltage.
The middle layer is a ground plane for reflecting impinging waves. 
The ground plane is connected to the patches on both sides of the top layer through four via-holes serving as ground, for design convenience.
The bottom layer contains direct current (DC) biasing lines that regulate the varactor diodes on the top layer.
Two via-holes connect the central patches to biasing lines. The copper thickness of the patches and biasing lines of the elements are all $ 35\, \mathrm{\mu m} $.
There are two kinds of substrates between the three layers, which are FR4 and F4B.
FR4 is a composite material composed of woven fiberglass cloth with an epoxy resin binder.
F4B is a series of high frequency materials, which are cheap in price and stable in quality.

\begin{figure}[t!]
	\centering
	\includegraphics[width=1.5in]{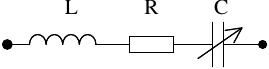}
	\caption{An equivalent circuit description of the varactor diode. It can be modeled as a series RLC circuit.}
	\label{fig:RLC}
\end{figure}

\begin{figure}[t!]
	\centering
	\includegraphics[width=\linewidth]{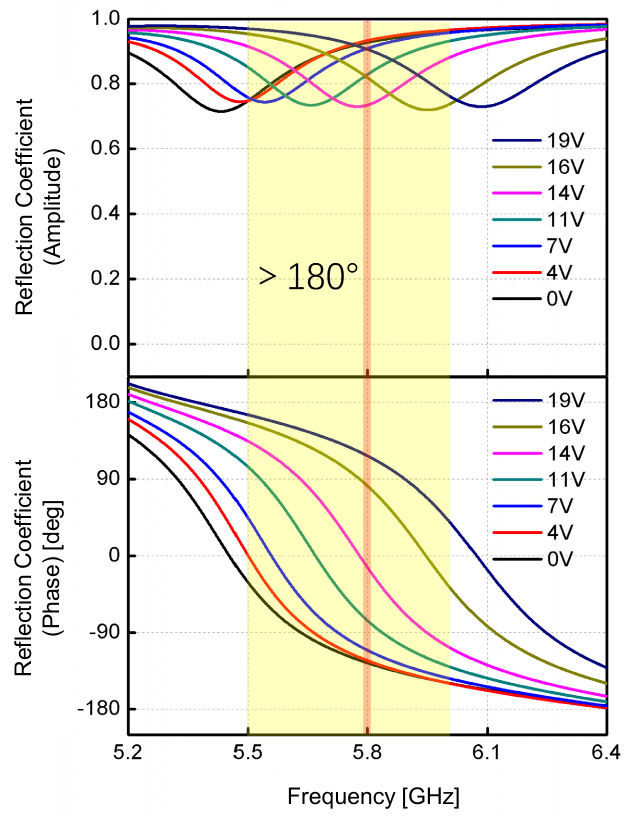}
	\caption{The simulated reflection amplitude and phase response of the RIS element under different voltages from 0  to 19 V. The yellow region represents the range where the maximum phase difference is larger than 180 degrees. The orange region will be used for communications in the field trials.}
	\label{fig:response}
\end{figure}

The properties of the proposed element were first evaluated with the commercial electromagnetic solver CST Microwave Studio 2019, using periodic boundary conditions to quantify the reflection spectra for different bias voltages.
The varactor diodes integrated into the element are Skyworks SMV2019-079LF, which can be modeled as a series Resistor-Inductor-Capacitor (RLC) circuit (i.e., it consists of a resistance, a capacitance, and an inductance connected in series) as shown in \figurename{~\ref{fig:RLC}}. The equivalent circuit component parameters are shown in Table {\ref{tab:Dparameters}}, where ``VR" denotes the Reverse Voltage applied on the diode.
The varactor diode is modeled by a lumped port in the simulation.
The amplitude and phase responses of the reflection under different bias voltages are shown in \figurename{~\ref{fig:response}}.
The observed behavior is aligned with the canonical case described in \cite{RIS_SPMAG}, where different voltages result in widely different phase shifts over the considered frequency range. 
For any given frequency within the yellow-marked band from 5.5 GHz to 6.0 GHz, varying the bias voltage from 0 to 19 V is sufficient to modify the phase-shift by at least 180 degrees.
The amplitude variations are substantially smaller within the considered band and the largest losses appear when the phase is close to zero.

\begin{table}[t!]
	\renewcommand{\arraystretch}{1.3}
	\caption{Equivalent Circuit Parameters of Varactor Diode\cite{Skyworks} }
	\label{tab:Dparameters}
	\centering
	\begin{tabular}{|cccc|}
		\hline\hline
		\textbf{VR (V)} & \textbf{C (pF)} & \textbf{R (ohm)} & \textbf{L (nH)} \\ \hline
		       0        &      2.31       &       4.51       &      0.70       \\ \hline
		      -4        &      0.84       &       4.04       &      0.70       \\ \hline
		      -7        &      0.55       &       3.66       &      0.70       \\ \hline
		      -11       &      0.38       &       3.18       &      0.70       \\ \hline
		      -14       &      0.31       &       2.86       &      0.70       \\ \hline
		      -16       &      0.27       &       2.65       &      0.70       \\ \hline
		      -19       &      0.24       &       2.38       &      0.70       \\ \hline\hline
	\end{tabular}
\end{table}

We have designed a 1-bit prototype where the two states have a phase difference of around 180 degrees. Note that this granularity is sufficient to configure a set of elements so their reflected signals are (partial) coherently combined at a location of choice \cite{Wu2020b}.
The orange region in the middle of \figurename{~\ref{fig:response}} represents a 20 MHz bandwidth around the central frequency of 5.8 GHz.
We can conclude from the figure that the considered element structure can achieve a 180 degree phase difference in this region, for example, using 7 V and 16 V as bias voltages, for which the phase shifts are around $-90$ and $+90$ degrees, respectively.
Moreover, the reflection coefficients are approximately constant within the considered bandwidth, thus the RIS element has a frequency-flat response.

\subsection{RIS Board Design}

The RIS board consists of a multitude of RIS elements and their controllers.
In \cite{7510962}, an RIS that can provide continuous phase shift variations for each element was proposed.
Its micro-controllers generate Pulse-Width Modulation (PWM) signals that are low-pass filtered to create bias voltages that can be continuously varied from 0 to 5 V. 
In this method, the PWM signal controlling each element needs to be generated independently, and a corresponding low-pass filter circuit is needed. Thus, the cost and hardware complexity will increase dramatically when  building a large-scale RIS.
A high phase-shift resolution might be useful when using an RIS to cancel interference in multi-user scenarios \cite{Wu2019}, but the goal of our field trials is for the RIS to reflect the incident waves as a beam pointing at the receiver.
In such scenarios, low-precision phase configuration is sufficient to avoid that signals reflected by the individual elements are interfering destructively \cite{7480359}. The power loss compared to a high-precision configuration is only a few dB \cite{Wu2020b}.
To limit the complexity of the prototype, we used two voltages to represent the phase states of $-\pi/2$ and $\pi/2$. 

\begin{figure}[t!]
	\centering
	\includegraphics[width=\linewidth]{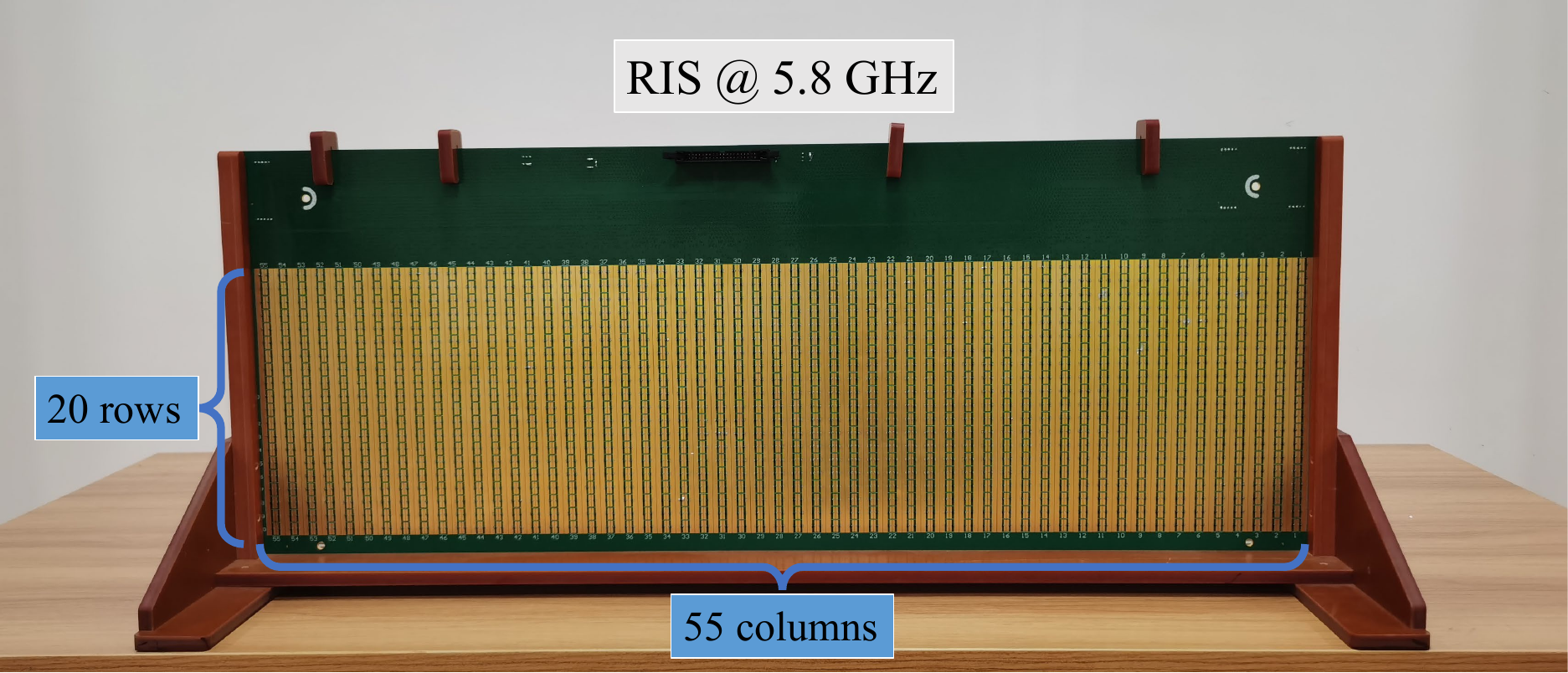}
	\caption{The front view of the fabricated $ 55 \times 20 $ RIS, each element capable of 1-bit phase shifting. The size is $ 80.08 \, \mathrm{cm} \times 31.30 \, \mathrm{cm} $.}
	\label{fig:RISfrontview}
\end{figure}

Based on the proposed element structure, we designed and fabricated an RIS board. It has the form of a Uniform Planar Array (UPA) with a $ 55 \times 20 $ element grid.
The front view of the RIS board is shown in \figurename{~\ref{fig:RISfrontview}}. The size of the fabricated board is $ 80.08 \, \mathrm{cm} \times 31.30 \, \mathrm{cm} $, which also include the control circuit, i.e., the green part that does not have the capability of regulating the impinging radio waves. The size of the RIS part is $78.65 \, \mathrm{cm} \times 20.54 \, \mathrm{cm}$. 

\begin{figure*}[t!]
	\centering
	\includegraphics[width=\linewidth]{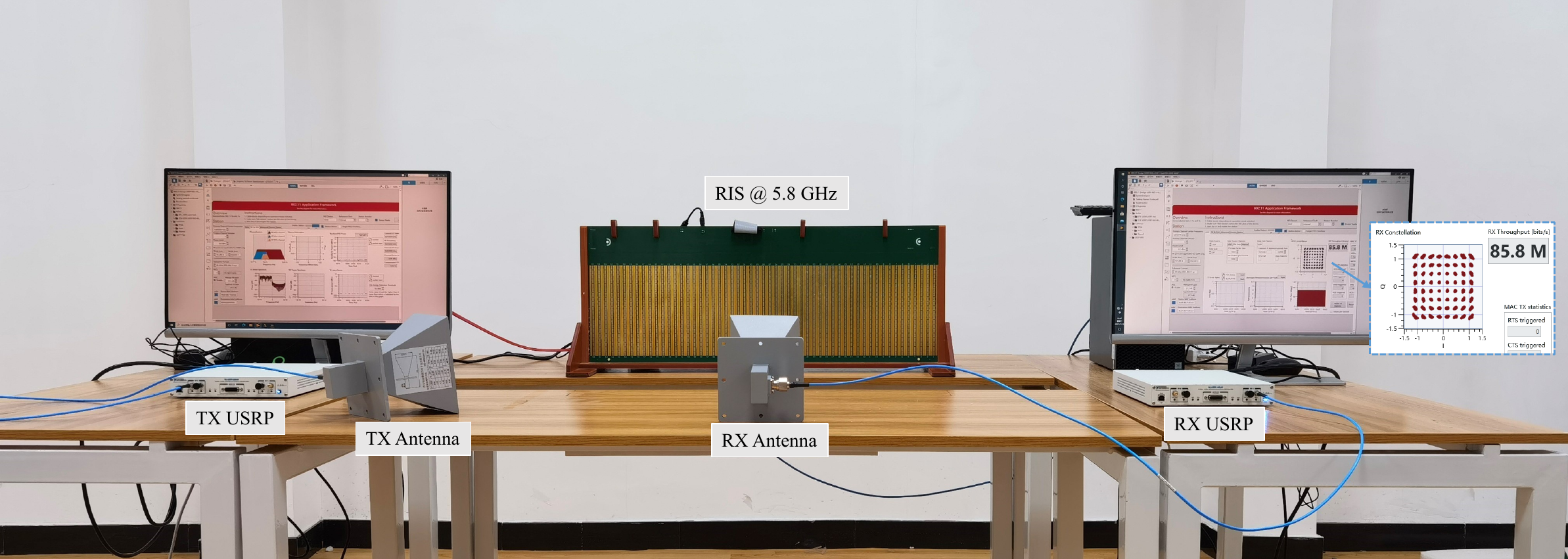}
	\caption{The prototype of the RIS-aided wireless communication system.} \vspace{-3mm}
	\label{fig:prototype}
\end{figure*}

In the typical wireless applications, the users can move around freely in the azimuth plane, but do not have much freedom to move in the vertical direction. As a result, the RIS requires a high level of flexibility when controlling beam directions in the azimuth plane, while a lower accuracy is required in the vertical domain.
We exploited this property to further simplify the implementation. More precisely, every five elements in each column form a group that shares the same bias voltage and, therefore, the same reflection coefficient. Hence, the total number of controllable bias signals is 220. 
The control system is composed of an FPGA and 28 serial-in and parallel-out 8-bit shift registers, which generate parallel discrete bias voltages for the RIS elements. 
After that, the voltage is adjusted by a voltage level-shifter so that the phase difference at the working frequency of 5.8 GHz is 180 degrees when the digital state changes from logic ``0'' to logic ``1''. 
The bias voltages are applied to each element to adjust the junction capacitance of the varactor diodes. 

\section{RIS-aided Wireless Communication System}\label{sec:system}

In this section, we describe the prototype of the RIS-aided wireless communication system shown in \figurename{~\ref{fig:prototype}}, which is composed of Personal Computers (PCs), universal software radio peripherals (USRPs), an RIS, an FPGA-based master control board, and the real-time RIS-UE feedback module.

The host computer, a USRP, and an antenna are the main components of the transmitter.
The receiver consists of the same main components. These transceivers emulate an Access Point (AP) or base station (BS) and a user equipment (UE), respectively. The transmitter sends a video stream using a single passive antenna. The data is transmitted through the User Datagram Protocol (UDP). The receiver decodes the data stream and plays the video on the host computer. 

\begin{figure}[t!]
	\centering
	\includegraphics[width=\linewidth]{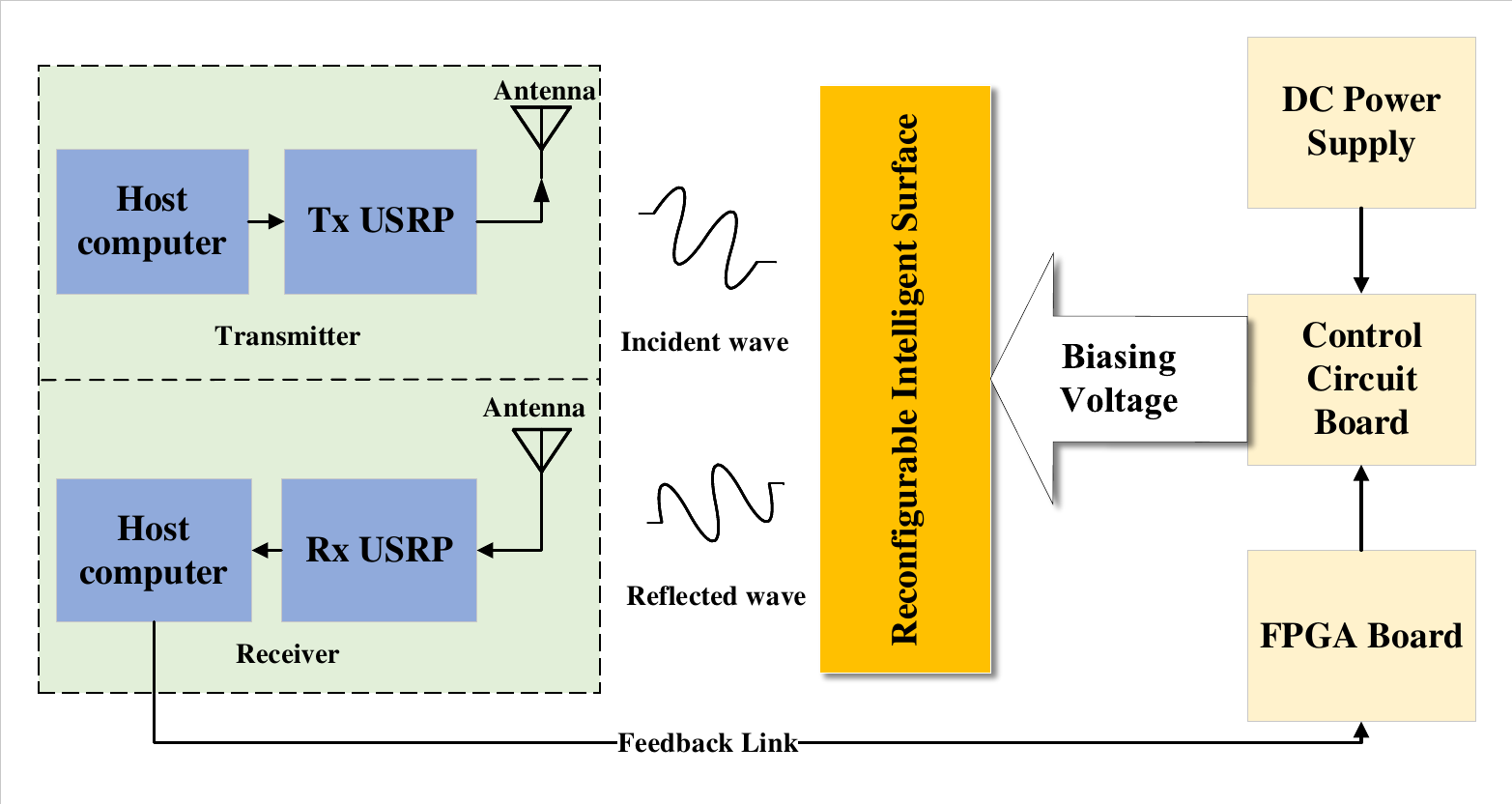}
	\caption{System diagram for the RIS-aided communication system.}
	\label{fig:SystemDiagram}
\end{figure}

\figurename~\ref{fig:SystemDiagram} shows the hardware block diagram for the prototype of an RIS-aided wireless communication system.
Table {\ref{tab:Hardware}} summarizes the details of the hardware modules.
In the prototype system, the goal of the RIS is to serve as a smart reflector that ``reflects'' the impinging wireless signal as a beam towards the receiver.
To support real-time adjustment of the reflection coefficients (e.g., to adapt the RIS to time-varying channel conditions or switch between serving different UEs), we introduce a feedback link \cite{Bjornson2020} that connects the RIS board and the receiver with serial ports. 
The receiver feeds back the quality of the current channel (i.e., the received signal power), to the FPGA master control board. The feedback can be used to measure the effectiveness of the applied reflection coefficients. The next section will describe how we utilized this feedback for beamforming design.

\begin{table}[t!]
	\renewcommand{\arraystretch}{1.3}
	\caption{Detailed Information of Hardware Modules}
	\label{tab:Hardware}
	\centering
	\begin{tabular}{|c|c|}
		\hline\hline
		  \textbf{Name}    &            \textbf{Description}            \\ \hline
		Central Controller & Computer with Intel Core i7-9700 processor \\ \hline
		       SDR         &               NI USRP-2954R                \\ \hline
		   FPGA Module     &  Development board with Xilinx Zynq 7100   \\ \hline
		   Antenna    Gain &          17.1 dBi gain @ 5.8 GHz           \\ \hline
		   {Antenna  Aperture } &         { $169 \, \mathrm{mm} \times 119 \, \mathrm{mm}$           }\\ \hline
		   {Antenna    Beamwidth} &          {30$^\circ$}           \\ \hline\hline
	\end{tabular}
\end{table}

\begin{figure}[t!]
	\centering
	\includegraphics[width=\linewidth]{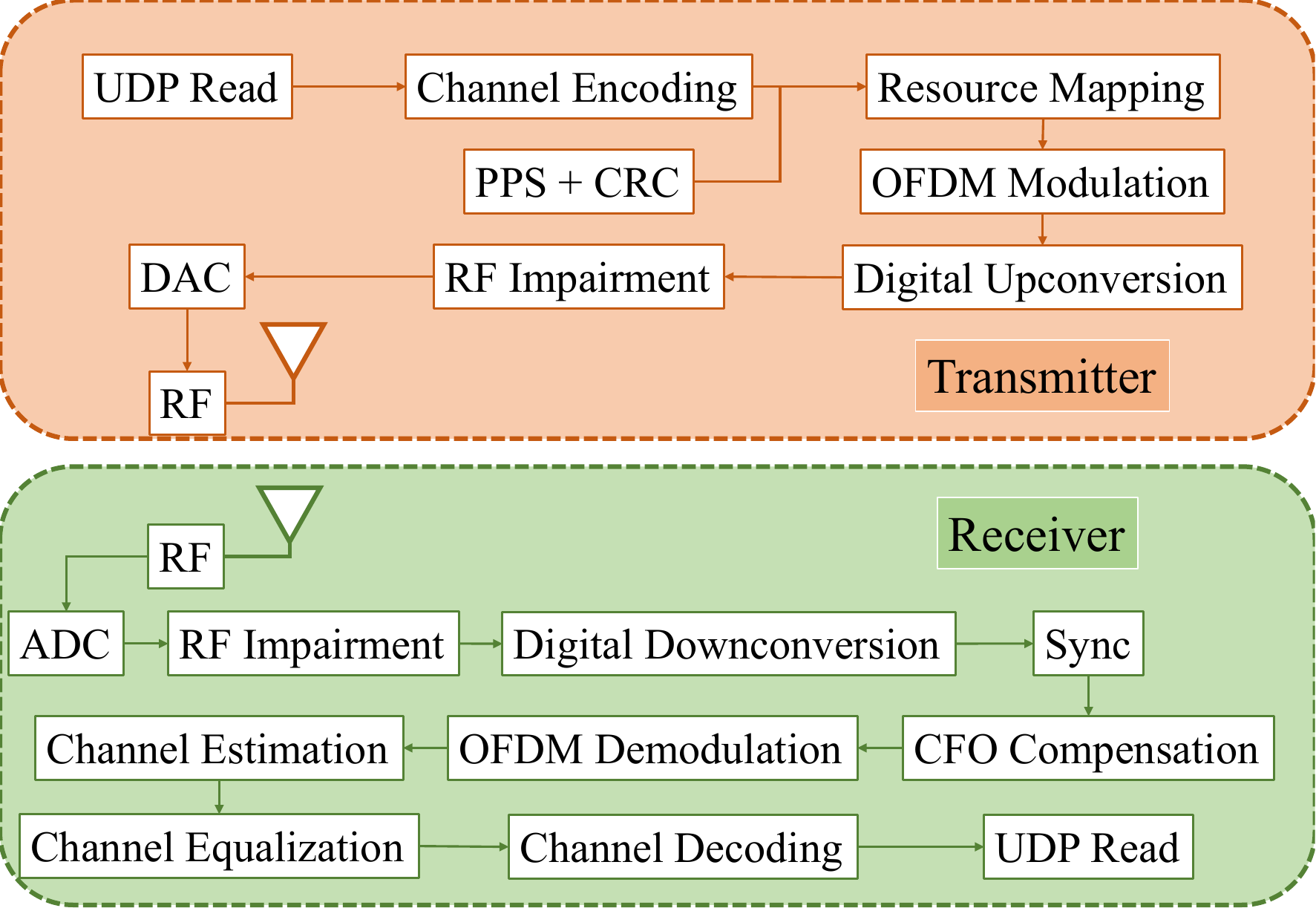}
	\caption{The detail framework and the signal processing flow in USRPs.}
	\label{fig:SignalFlow}
\end{figure}

The following part introduces the software framework of our communication system.
The signal processing flow is shown in \figurename{~\ref{fig:SignalFlow}}.
First, our transmitter loads data provided by an external application, from a UDP socket.
After the payload data is encoded, the samples are mapped to the time-frequency resources.
Afterwards, the Orthogonal Frequency Division Multiplexing (OFDM) modulation is applied, which converts the frequency-domain parallel signals to time-domain serial signals. The bandwidth and subcarrier spacing are 20 MHz and 312.5 kHz, respectively, in our experiments. 
The signal is then upconverted to the carrier frequency and transmitted through the transmit antenna. 

The receiver module performs the opposite functions as the transmitter.
Synchronization and Carrier Frequency Offset (CFO) compensation are achieved by continuous measurement of both autocorrelations and cross correlations of the pilot signals.
Note that the existing {IEEE 802.11ac} WiFi standards are adopted in this prototype system, and the over-the-air signaling of WiFi is kept intact.
In other words, the integration of the RIS technology into the existing communication networks can be made seamless without  modifications of the wireless communication protocols.                                    

\section{RIS-aided Beamforming Method}\label{sec:codebook}

The RIS will scatter or reflect the incident wave in a way determined by its configuration. In a scenario with a single-antenna transmitter and single-antenna receiver, as shown in \figurename~\ref{fig:SystemDiagram}, the communication performance is determined by the Signal-to-Noise Ratio (SNR). Hence, the RIS can aid the communication system by optimizing its reflection coefficients to maximize the received signal power. This effectively means that the RIS elements should make their individually reflected signals reach the receiver in phase (up to the granularity set by the hardware). If there is an LoS channel from the RIS to the receiver, the optimization will essentially result in a reflected beam in the angular direction leading to the receiver.

The RIS beamforming configuration problem has been analyzed mathematically in a series of previous works\cite{Liu2020matrix, Wei2020Parallel}; we refer to \cite{9326394} for a recent survey.
In previous literature, the RIS configuration is generally divided into two disjoint subproblems: 1) selecting an SNR-maximizing configuration when having full Channel State Information (CSI) \cite{Huang2018a,Wu2019}; 2) Acquiring CSI from explicit pilot signals  \cite{Taha2019a,Zheng2020,wang2020}.
The latter part is particularly challenging since the RIS (normally) does not have any RF chains, which prohibits it from directly obtaining CSI regarding the BS-RIS or RIS-UE channels using traditional pilot-based channel estimation methods \cite{massivemimobook}.
Full-dimensional CSI acquisition requires the pilot signaling to grow linearly with the number of RIS elements \cite{Bjornson2020}.
To the best of our knowledge, the existing works rely on high-precision configurations where an RIS with $L$ elements can switch between $L$ mutually orthogonal configurations. 
In summary, none of the existing algorithms was deemed to be implementable in our practical setup, where CSI acquisition and RIS configuration must be done jointly, and where there is only a low-bit resolution per element.
Hence, we had to develop a new algorithm that could be utilized in the prototype system.

\subsection{Codebook-based RIS Beamforming}

Since the fabricated RIS has the form of a UPA and is intended for scenarios with dominant LoS paths, we can exploit the geometry to develop a low-complexity method to configure the RIS.
We will take inspiration from the codebook-based beamforming for UPAs.
In traditional MIMO communications, a two-dimensional (2D) Discrete Fourier Transform (DFT) codebook can be used to specify a set of angular transmit beamforming directions when using a UPA.
While in RIS-aided communications, the angle of the reflected beam depends on both the incident angle and the RIS configurations, we will show that for a given incident angle and a desired reflecting angle, the structure of the optimal RIS phase shifts is still in line with the 2D-DFT codebook.

\begin{figure}[t!]
	\centering
	\includegraphics[width=0.8\linewidth]{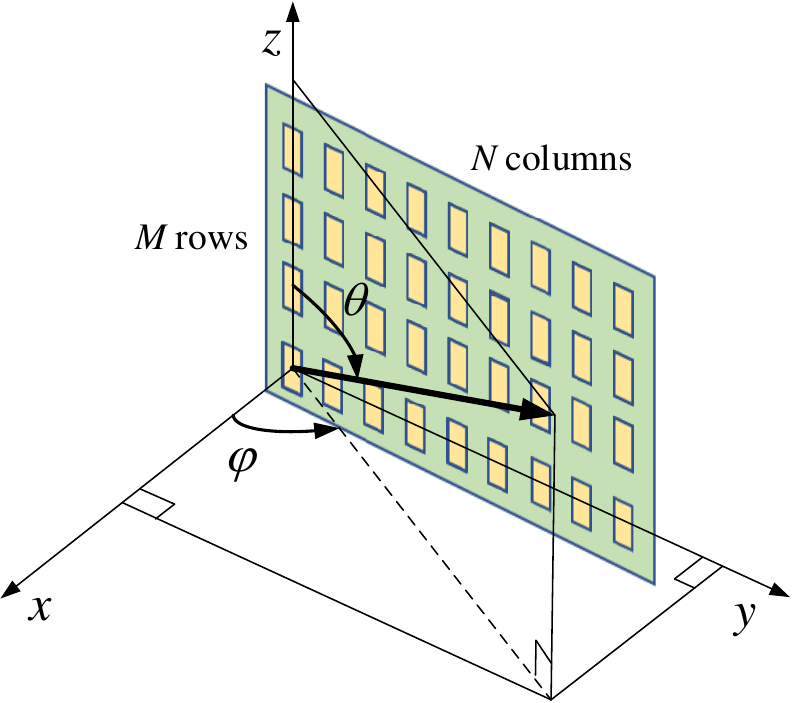}
	\caption{The coordinate system describing the RIS with $M$ rows and $N$ columns.}
	\label{fig:UPA_model}
\end{figure}

As shown in \figurename~\ref{fig:UPA_model}, the RIS consists of a total of $ L=M  N $ elements, where ${M}$  represents the number of RIS elements in the \textit{z}-axis direction (i.e., the number of rows), and ${N}$ represents the number of RIS elements in the \textit{y}-axis direction (i.e., the number of columns).
The elements are deployed on a rectangular grid with $ d_y $ representing the spacing between the \textit{y}-axis elements, and $ d_z $ represents the spacing between the \textit{z}-axis elements.
As in most metamaterial designs, the vertical and horizontal spacings are smaller than half the wavelength, i.e.,  $ d_x\le \lambda /2  $ and $ d_y \le \lambda / 2 $, where $ \lambda $ is the wavelength of the intended signal. 

The classical far-field approximation is known to be accurate even for large but finite surfaces \cite{BS20} and will be utilized herein.
The incident wave will consist of a superposition of multiple plane waves from different angles. We focus on one of them.
The first reflecting element is at the origin. 
The zenith angle and azimuth angle to the transmitter, as seen from the origin, are $ \theta $ and $ \varphi $ respectively. 
For such an incident wave, the array response vector $ \mathbf{a}(\theta, \varphi) \in \mathbb{C}^{L \times 1} $
of the UPA is \cite{3gpp:38.901,yin2020JSAC}:
\begin{equation}
	\mathbf{a}(\theta, \varphi)=\mathbf{a}_{y}(\theta, \varphi) \otimes \mathbf{a}_{z}(\theta), \label{eq:SteeringVector}
\end{equation}
where $ \otimes $ denotes the Kronecker product. $ \mathbf{a}_{y}(\theta, \varphi) $ and $ \mathbf{a}_{z}(\theta, \varphi) $ represent the array response vectors of a uniform linear array along the \textit{y}-axis and the \textit{z}-axis respectively:
\begin{align}
	& \mathbf{a}_{y}(\theta, \varphi) = \nonumber \\
	& \quad \left[1, e^{\frac{-j 2 \pi}{\lambda} d_{y} \, \sin \theta \sin \varphi },
	\cdots,
	e^{\frac{-j 2 \pi}{\lambda} d_{y} \,(N-1) \, \sin \theta \sin \varphi}
	\right]^\mathrm{T}, \\
	& \mathbf{a}_{z}(\theta) =
	\left[1,
	e^{\frac{-j 2 \pi}{\lambda} d_{z} \cdot\cos \theta},
	\cdots,
	e^{\frac{-j 2 \pi}{\lambda} d_{z} \cdot(M-1) \, \cos \theta}
	\right]^\mathrm{T}.
\end{align}

Note that in this model, we order the RIS elements column by column, starting from the origin.\footnote{If they are ordered row by row, we simply need to switch the order of the Kronecker product in (\ref{eq:SteeringVector}).}
In the considered multi-carrier system, the RIS-UE channel matrix is denoted as 
\begin{align}
    {\bf{G}} = [{{\bf{g}}_1},{{\bf{g}}_2}, \cdots ,{{\bf{g}}_K}] \in \mathbb{C}^{L \times K},
\end{align}
where $K$ is the number of subcarriers and ${{\bf{g}}_k}$ is the channel vector at the $k$-th subcarrier, for $k = 1, \dots, K$. 
This channel matrix can be expressed as \cite{3gpp:38.901,yin2020JSAC}
\begin{align} \label{eq:G-matrix}
    {\bf{G}} = {\alpha _g}{\bf{a}}({\theta _g},{\varphi _g}){ {{\bf{b}}^\mathrm{T}({\tau _g})} },
\end{align}
where $\alpha _g \in \mathbb{C}$ is the complex path gain and ${\bf{a}}({\theta _g},{\varphi _g})$ is the array response vector in \eqref{eq:SteeringVector} with 
$\theta_g$ and $\varphi_g$ being the zenith angle and azimuth angle of the UE.
The delay response vector ${\bf{b}}({\tau _g}) \in \mathbb{C}^{K \times 1}$ is defined as
\begin{align} 
{\bf{b}}({\tau _g}) = \left[ {\begin{array}{*{20}{c}}
{{e^{ - j2\pi {f_1}{\tau _g}}}},& \cdots ,&{{e^{ - j2\pi {f_K}{\tau _g}}}}
\end{array}} \right]^\mathrm{T},
\end{align}
where $\tau _g$ is the path delay and $f_k$ is the frequency of the $k$-th subcarrier. 
If there are multiple paths, then ${\bf{G}}$ is computed as the summation over multiple terms of the type ${\alpha _g}{\bf{a}}({\theta _g},{\varphi _g}){ {{\bf{b}}^\mathrm{T}({\tau _g})} }$ with different values of $\alpha_g,\theta_g,\varphi_g$, and $\tau_g$ \cite{yin2020JSAC}.

Similarly, the multi-carrier BS-RIS channel $ \mathbf{H} \in \mathbb{C}^{L \times K}$ is denoted as
\begin{align}
    {\bf{H}} = [{{\bf{h}}_1},{{\bf{h}}_2}, \cdots ,{{\bf{h}}_K}] \in \mathbb{C}^{L \times K},
\end{align}
where ${{\bf{h}}_k}$ is the channel at the $k$-th subcarrier. This channel matrix can be computed as
\begin{align} \label{eq:H-matrix}
    {\bf{H}} = {\alpha _h}{\bf{a}}({\theta _h},{\varphi _h}){ {{\bf{b}}^\mathrm{T}({\tau _h})} },
\end{align}
where $\theta_h$ and $\varphi_h$ denote the zenith angle and azimuth angle of the BS, $\alpha _h$ is the complex path gain, and $\tau _h$ is the BS-RIS path delay. If there are multiple paths, then \eqref{eq:H-matrix} can be generalized as a summation over these terms.

In RIS-aided communications without a direct path, the received signal $y_k \in \mathbb{C}$ at the $k$-th subcarrier can be expressed as \cite{Zheng2020,RIS_SPMAG}
\begin{equation}
	y_k = (\mathbf{g}_k \odot \mathbf{h}_k)^\mathrm{T} \boldsymbol{\omega} x_k
 + n_k,
\end{equation}
where $x_k \in \mathbb{C}$ is the transmitted signal, $\odot$ denotes the Hadamard (element-wise) product, and $ n_k \sim \mathcal{CN}({0, \sigma_n^2})$ is the white Gaussian noise with variance $\sigma_n^2$.
The vector $\boldsymbol{\omega} \in \mathbb{C}^{L \times 1}$ contains the reflection coefficients of the RIS.
The SNR at the $k$-th subcarrier is proportional to the end-to-end power gain $| (\mathbf{g}_k \odot \mathbf{h}_k)^\mathrm{T} \boldsymbol{\omega}|^2$ and can be tuned by the selection of $\boldsymbol{\omega}$.

The same $\boldsymbol{\omega}$ must be used on every subcarrier, which could lead to complicated RIS optimization problems \cite{Zheng2020}. We focus on scenarios with a single dominant path and utilize \eqref{eq:G-matrix} and \eqref{eq:H-matrix} to simplify the end-to-end power gain as
\begin{equation}
    | (\mathbf{g}_k \odot \mathbf{h}_k)^\mathrm{T} \boldsymbol{\omega}|^2 = |\alpha_g |^2 |\alpha_h|^2 \left| ({\bf{a}}({\theta _g},{\varphi _g}) \odot {\bf{a}}({\theta _h},{\varphi _h}))^\mathrm{T} \boldsymbol{\omega}\right|^2,
\end{equation}
which is independent of the subcarrier index.

Since our reflecting elements are controlled by a single bias voltage, it does not have the freedom of independent phase and amplitude modifications. Therefore, we only consider the phase-shift functionality of each RIS element. If a continuous phase-shift optimization is possible, then the optimal phase-shift vector is given as \cite{Wu2019,Zheng2020}
\begin{align}
	\boldsymbol{\omega}_o &= \underset{|[\boldsymbol{\omega}]_{i}| = 1}{\arg \max }  \, \left| ({\bf{a}}({\theta _g},{\varphi _g}) \odot {\bf{a}}({\theta _h},{\varphi _h}))^\mathrm{T} \boldsymbol{\omega}\right|^2 \\ 
	&= \left(\mathbf{a}\left(\theta_{g}, \varphi_{g}\right) \odot \mathbf{a}\left(\theta_{h}, \varphi_{h}\right)\right)^{*}, \label{eq:opt-omega}
\end{align}
where $[\boldsymbol{\omega}]_{i}$ denotes the $i$-th entry of $\boldsymbol{\omega}$ and $*$ denotes the complex conjugate. The second equality follows from the Cauchy-Schwartz inequality.
The optimal solution selects the phase-shifts so the reflected signals from all RIS elements arrive in phase at the receiver.
By utilizing \eqref{eq:SteeringVector}, we can rewrite \eqref{eq:opt-omega} as
\begin{align}
\boldsymbol{\omega}_o  =  {{{\left( {{{\bf{a}}^{(d)}}\left( {{\theta _g},{\varphi _g},{\theta _h},{\varphi _h}} \right)} \right)}^*}}, \label{eq:desired-configuration}
\end{align}
where 
\begin{align}
 & {\bf{a}}^{(d)}\left( {{\theta _g},{\varphi _g},{\theta _h},{\varphi _h}} \right) = \nonumber \\ 
 & {\bf{a}}_y^{(d)}({\theta _g},{\varphi _g},{\theta _h},{\varphi _h})  
  \otimes  {\bf{a}}_z^{(d)}({\theta _g},{\theta _h}), \label{Eq:ad}
\end{align}
with 
\begin{align}
& {\bf{a}}_y^{(d)}({\theta _g},{\varphi _g},{\theta _h},{\varphi _h}) = \left[ 1,{e^{\frac{{ - j2\pi }}{\lambda }{d_y}\left( {\sin {\theta _h}\sin {\varphi _h} + \sin {\theta _g}\sin {\varphi _g}} \right)}}, \right. \nonumber \\
& \qquad  \left. \cdots ,{e^{\frac{{ - j2\pi }}{\lambda }{d_y}(N - 1)\left( {\sin {\theta _h}\sin {\varphi _h} + \sin {\theta _g}\sin {\varphi _g}} \right)}} \right]^\mathrm{T}, \\
& {\bf{a}}_z^{(d)}({\theta _g},{\theta _h}) = \left[ 1,{e^{\frac{{ - j2\pi }}{\lambda }{d_z}\left( \cos {\theta _h} + \cos {\theta _g} \right)}}, \right. \nonumber \\
& \qquad \left. \cdots ,{e^{\frac{{ - j2\pi }}{\lambda }{d_z}(M - 1)\left( {\cos {\theta _h} + \cos {\theta _g}} \right)}} \right]^\mathrm{T}.
\end{align}

The solution in \eqref{Eq:ad} has a structure that can be exploited in calculating the reflection coefficients. 
Note that both ${\bf{a}}_y^{(d)}({\theta _g},{\varphi _g},{\theta _h},{\varphi _h})$ and ${\bf{a}}_z^{(d)}({\theta _g},{\theta _h})$ have the structure $[1,\xi,\xi^2,\ldots]^\mathrm{T}$, but with different complex exponentials $\xi$ and lengths.
With the increase in the number of elements, such vectors can be well approximated by the columns of the DFT matrix \cite{yin:13,adhikary:13}. 
We denote the DFT matrix of size $ K \times K $ as
\begin{small}
\begin{equation}
\mathcal{F}(K)=  \left[\begin{array}{ccccc}
1 & 1 & 1 & \cdots & 1 \\
1 & \xi & \xi^{2} & \cdots & \xi^{K-1} \\
1 & \xi^{2} & \xi^{2 \cdot 2} & \cdots & \xi^{2(K-1)} \\
\vdots & \vdots & \vdots & \ddots & \vdots \\
1 & \xi^{K-1} & \xi^{2(K-1)} & \cdots & \xi^{(K-1)(K-1)}
\end{array}\right],
\end{equation}
\end{small}
where $\xi=e^{-j(2 \pi / K)}$.

For the considered UPA architecture, we can build a beamforming codebook for the RIS as $ \mathcal{F}(M) \otimes \mathcal{F}(N) $. 
Therefore, each column of $ \mathcal{F}(M) \otimes \mathcal{F}(N) $ forms a codeword representing the reflection of an incident beam in a certain beam direction. When $M$ and $N$ are large, the codewords will closely approximate all RIS configurations for which an incident plane wave is beamformed in a distinct angular direction.

In our prototype, the phase shifts are quantized to being either $-\pi/2$ or $+\pi/2$, thus the above-described solution cannot be utilized. Hence, the corresponding problem is 
\begin{align}
	\hat{\boldsymbol{\omega}}_o &= \underset{[\boldsymbol{\omega}]_{i} \in \{ e^{-j \pi/2},e^{j \pi/2} \}}{\arg \max }  \, \left| ({\bf{a}}({\theta _g},{\varphi _g}) \odot {\bf{a}}({\theta _h},{\varphi _h}))^\mathrm{T} \boldsymbol{\omega}\right|^2 \\ 
	&= \underset{[\boldsymbol{\omega}]_{i} \in \{ e^{-j \pi/2},e^{j \pi/2} \}}{\arg \max }  \, \left| \boldsymbol{\omega}_o^\mathrm{H} \boldsymbol{\omega}\right|^2,
	\label{eq:opt-omega-discrete}
\end{align}
that finds the quantized vector that gives the maximum inner product with the ideal solution in \eqref{eq:opt-omega}. One solution is obtained by making sure all terms in the inner product in \eqref{eq:opt-omega-discrete} have a positive real part:
\begin{equation}
     [\hat{\boldsymbol{\omega}}_o]_i = \begin{cases}
     e^{j \pi/2}, & \textrm{if } \arg( [{\boldsymbol{\omega}}_o]_i) \in [-\pi,0) \\
     e^{-j \pi/2}, & \textrm{if }  \arg( [{\boldsymbol{\omega}}_o]_i) \in [0,\pi).
     \end{cases}
 \end{equation}
In practice, when the angular information is not directly available, one may search through the codebook and find the codeword that gives the maximum received power. Although the codebook is designed based on far-field array response vectors, only a minor loss in beamforming gain is expected when operating the system at distances shorter than the Fraunhofer distance. We recall from \cite{BS20} that the main near-field effects only occurs when the propagation distance is proportional to the RIS size (in 80 cm in our prototype).

\subsection{Proposed Fast Beamforming Algorithm}

Most previous algorithms for channel estimation and RIS configuration treat these problems separately, consider arbitrary channel structure, and assume infinite phase-shift precision.
However, recent works have shown that the most compelling advantage of the RIS technology exists when the transmitter and receiver have a single dominant path to the RIS, and the RIS is configured based on these paths \cite{Zheng2020,RIS_SPMAG}. This insight is utilized in our prototype to greatly reduce the training/feedback overhead when configuring the surface. We will now describe the proposed RIS beamforming algorithm, which will be evaluated experimentally.

When there are dominant angular paths, the desired configuration
will be of the form in \eqref{eq:desired-configuration}, which is approximately a column of $ \mathcal{F}(M) \otimes \mathcal{F}(N) $. When quantizing the elements of the codewords to be either $e^{-j \pi/2}$ or $e^{j \pi/2}$, many adjacent elements will have the same phase shift. We will utilize this property to develop an algorithm that gradually changes the phase shifts on a column-by-column or row-by-row basis to gradually increase the SNR of the end-to-end link.

The basic idea of our proposed greedy algorithm is to invert the phases of a certain row or column of the RIS. If this configuration state increases the power of the received signal, compared to the previous state, the new state is utilized. The algorithm is enabled by our UE-RIS feedback module. 
The details are given in Algorithm \ref{alg:GreedyAlgorithm}. Note that the reflection coefficients at time index $t$ are now denoted by an $M \times N$ complex matrix $\mathbf{R}_t$ for ease of exposition. The vector $\boldsymbol{\omega}$ is the vectorization of this matrix.
The $(m,n)$-th element in $\mathbf{R}_t$ is the reflection coefficient of the element located at the $m$-th row and $n$-th column of the RIS. $\mathbf{R}_t$ can also be written as 
\begin{align}
{{\mathbf{R}}_t} = \left[{{\mathbf{r}}_{t,1}},{{\mathbf{r}}_{t,2}}, \cdots ,{{\mathbf{r}}_{t,N}} \right] =  {\left[{{\mathbf{s}}_{t,1}},{{\mathbf{s}}_{t,2}}, \cdots ,{{\mathbf{s}}_{t,M}}\right]^\mathrm{T}},
\end{align}
where ${\mathbf{r}}_{t,n} \in \mathbb{C}^{M \times 1}$, $n = 1, \dots, N$ are the coefficients of the $n$-th column and ${\mathbf{s}}_{t,m} \in \mathbb{C}^{N \times 1}$, $m = 1, \dots, M$ are the coefficients of the $m$-th row. 
We can switch the states of a column by changing its sign, since 
$e^{-j \pi/2}=-j$ and $e^{j \pi/2}=j$.
We denote the average power of the received signal at time index $t$ as $p_t$, which is computed over the whole signal bandwidth in order to reduce the influence of noise.

\begin{algorithm}
	\caption{Greedy Fast Beamforming Algorithm}
	\label{alg:GreedyAlgorithm}
	\begin{algorithmic}
		\STATE \textbf{Input: } The feedback of RX signal power $p_t$.
		\STATE \textbf{Output: } The reflection coefficients matrix.
		\STATE Initialize a reflection coefficients matrix $ \mathbf{R}_0\in \mathbb{C}^{M \times N }$;
		\STATE Receive initial feedback of the RX signal power $ p_0 $;
		
		\STATE // Horizontal search;
		\FOR {each $n \in [1, N]$}
		\STATE  $ \mathbf{R}_{n} \leftarrow [\mathbf{r}_{n-1,1},\cdots,\mathbf{r}_{n-1,n-1},- \mathbf{r}_{n-1,n},$ 
		\STATE $\quad \quad \quad \quad \quad \quad \quad \quad \mathbf{r}_{n-1,n+1}\cdots,\mathbf{r}_{n-1,N}] $;
		\STATE Receive feedback $ p_{n} $ using configuration $ \mathbf{R}_{n}$;
		\IF {$ p_{n-1} \ge p_{n}$}
		\STATE $ \mathbf{R}_{n} \leftarrow \mathbf{R}_{n-1}$;
		\ENDIF
		\ENDFOR
		\RETURN{reflection coefficients matrix $ \mathbf{R}_{N}$;}
		
		\STATE // Vertical search;
		\STATE Denote $ \mathbf{R}_{N} =[\mathbf{s}_{N,1},\mathbf{s}_{N,2},\cdots,\mathbf{s}_{N,M}]^\mathrm{T}$;
		\FOR {each $m \in [1, M]$}
		\STATE  $\mathbf{R}_{N+m} \!\leftarrow\! [\mathbf{s}_{N+m-1,1},\cdots,\mathbf{s}_{N+m-1,m-1},- \mathbf{s}_{N+m-1,m},$ 
		\STATE $\quad \quad \quad \quad \quad \quad \quad \quad \mathbf{s}_{N+m-1,m+1},\cdots,\mathbf{s}_{N+m-1,M}]^\mathrm{T}$;
		\STATE Receive feedback $ p_{N+m} $ using configuration $\mathbf{R}_{N+m}$;
		\IF {$ p_{N+m-1} \ge p_{N+m}$}
		\STATE $ \mathbf{R}_{N+m} \leftarrow \mathbf{R}_{N+m-1}$;
		\ENDIF
		\ENDFOR
		\RETURN{reflection coefficient matrix $ \mathbf{R}_{M+N}$.}
	\end{algorithmic}
\end{algorithm}

Algorithm \ref{alg:GreedyAlgorithm} determines the reflection coefficients using $ M+N $ feedback iterations, which is much smaller than the $MN$ iterations required for estimating unstructured channels \cite{Bjornson2020}. The computational complexity order is $\mathcal{O}(M+N)$.
In addition, it does not introduce significant fluctuations of the received power, thus data transmission can be carried out simultaneously. The received power can be measured by the wideband SNR or Reference Signal Receiving Power (RSRP), or other similar metrics that reflect the channel quality. In particular, existing RSRP measurements can be utilized so that extra signaling is not needed. 
It is also possible to repeat Algorithm \ref{alg:GreedyAlgorithm} several times to refine the coefficients, and even keep it iterating during data transmission to track the UE's movement. We have observed from our experiments that ten iterations of Algorithm \ref{alg:GreedyAlgorithm} brings around 1 dB gain compared to the case of single iteration.
The algorithm can be initialized in different ways. We propose to start from a homogeneous configuration (i.e., all elements have the same state), which corresponds to the behavior of a conventional homogeneous reflective surface with the same absorption properties. In this way, the algorithm is guaranteed to outperform such a surface.

Even if the algorithm was designed particularly for scenarios with one dominant far-field path to the RIS and one dominant far-field path from the RIS, it can be applied under any circumstances (e.g., in near-field propagation) and will monotonically increase the SNR, thus it is also guaranteed to converge (if the channel conditions are static).
In the next section, we show the effectiveness of the proposed algorithm through experiments. 

\section{Experimental Results}\label{sec:results}

In this section, we present the results from several experiments that were made using our prototype to verify the performance of this proof-of-concept.
First, we test the electromagnetic reflection characteristics of the RIS.
Next, we verify the channel reciprocity, which is of particular value for communication theorists and engineers.
We then test the radiation pattern in an anechoic chamber to measure the beamforming function of our RIS.
Finally, we present and analyze the results from indoor and outdoor field trials.

\subsection{Bias Voltage Response Test}

\begin{figure}[t!]
	\centering
	\includegraphics[width=\linewidth]{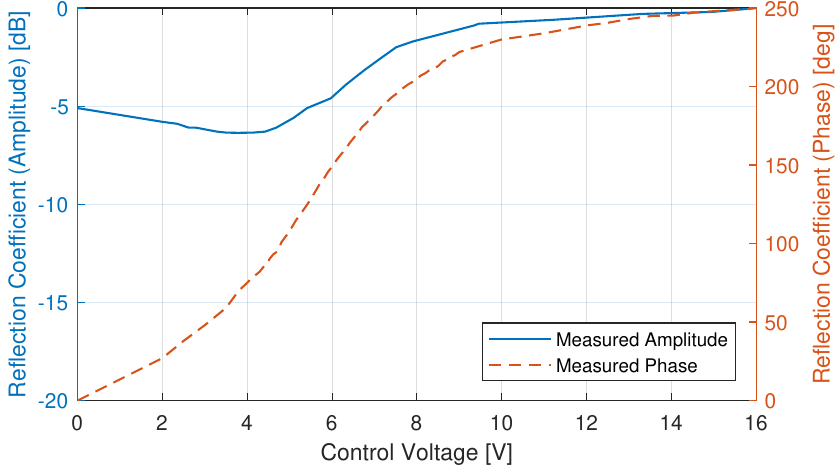}
	\caption{Measured relationship between the control voltage of RIS and the gain and phase of the reflection coefficient at 5.8 GHz. {The angles of incidence and reflection are both 45$^\circ$. }}
	\label{fig:gain_phase}
\end{figure}

Due to the tolerances of the PCB manufacturing process, the difference in the dielectric constant of the material, and the non-ideal characteristics of the varactor diode, the electromagnetic characteristics of a fabricated RIS are often different from the ideal simulated one \cite{dai2019wireless}.
To achieve precise control of the reflected signal, we measure the relationship between the reflection coefficients and the bias voltage of the varactor diode. 
We use horn antennas to transmit and receive signals that are reflected by the RIS.
Two antennas are placed 1 m away from the metasurface, and a sinusoidal signal at 5.8 GHz is used for the excitation.
As shown in \figurename{~\ref{fig:gain_phase}}, when the bias voltage gradually increases from 0 to 16 V, the experimental results show that the gain fluctuation is 6.5 dB and the range of phase shifts is 250 degrees. Note that the phases are normalized such that the phase shift is zero when the bias voltage is 0 V. The gain is also normalized and the maximum value is 0 dB. We observe from \figurename{~\ref{fig:gain_phase}} that the reflection behavior is well in line with the CST simulation results. 
However, the control voltages to generate each states of phase shift are not identical to the simulated ones. In order to achieve the desired reflection coefficients,  the control voltages need to be selected according to measurements. 
Hence, when designing a 1-bit RIS where each element can take two states having 180 degree phase difference, measurements must be made to identify the right bias voltages.
We find that when the bias voltages are 3.8 V and 16 V, the phase shifts are 70 degrees and 250 degrees, respectively, which leads to the desired phase difference of 180 degrees.

\begin{figure}[t!]
	\centering
	\includegraphics[width=\linewidth]{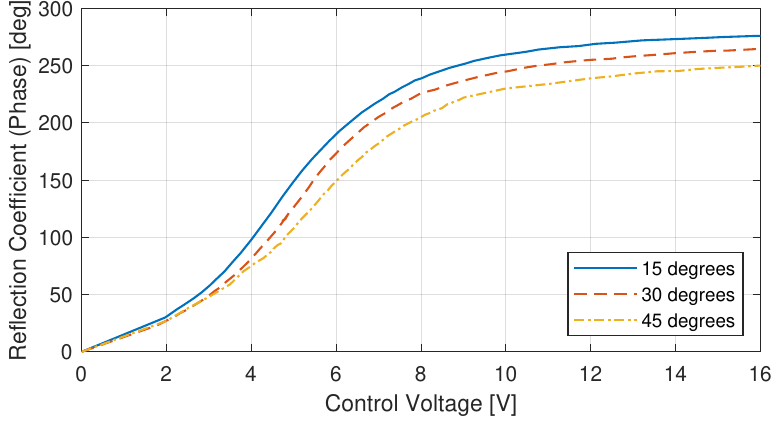}
	\caption{Measured relationship between the control voltage of the RIS and the phase of reflection coefficients with different incident angles at 5.8 GHz.}
	\label{fig:voltage_phase_all}
\end{figure}

We tested the voltage-phase response at different incident angles: $ 15^\circ $, $ 30^\circ $, and $ 45^\circ $ in the azimuth plane. The results are shown in \figurename{~\ref{fig:voltage_phase_all}}.
When the bias voltage increases from 0 to 16 V, a dynamic phase range of about $ 276^\circ $ can be achieved at $ 15^\circ $, $ 265^\circ $ can be achieved at $ 30^\circ $, and only $ 250^\circ $ can be achieved at $ 45^\circ $.
Importantly, the experimental results show that the response of the RIS elements is angle-dependent, thus one may not find a pair of bias voltages that will result in exactly 180 degrees phase difference for any incident angles. A larger incident angle results in smaller phase variations.
This observation is in line with the results in \cite{Tang2021}, where the phase variations due to different incident angles were even bigger. 
The likely physical explanation of this phenomenon is that the reflectance of the element depends on the incident angle \cite{Zhang2020a}.
This property has not been taken into consideration in the canonical system model used by the communication and signal processing literature \cite{9326394,RIS_SPMAG}. It is important to evolve the models since angle-dependence implies that different paths in a multipath scenario will be subject to different phase shifts, and that one cannot design a codebook of RIS configuration vectors that are orthogonal for all incident angles.

\subsection{Channel Reciprocity Test}

The time division duplex (TDD) mode is widely adopted in 5G mobile communication systems and WiFi. 
One feature is that the uplink (UL) and downlink (DL) channels are reciprocal, which is a great advantage when having many antennas on one side of the link, because one can then learn the channels in the direction where the channel estimation requires the least signaling. For example, to support an arbitrarily large number of base station antennas,  massive MIMO builds on the principle of estimating channels in the UL and then uses the estimates also for DL precoding \cite{massivemimobook}.

\begin{figure}[t!]
	\centering
	\includegraphics[width=\linewidth]{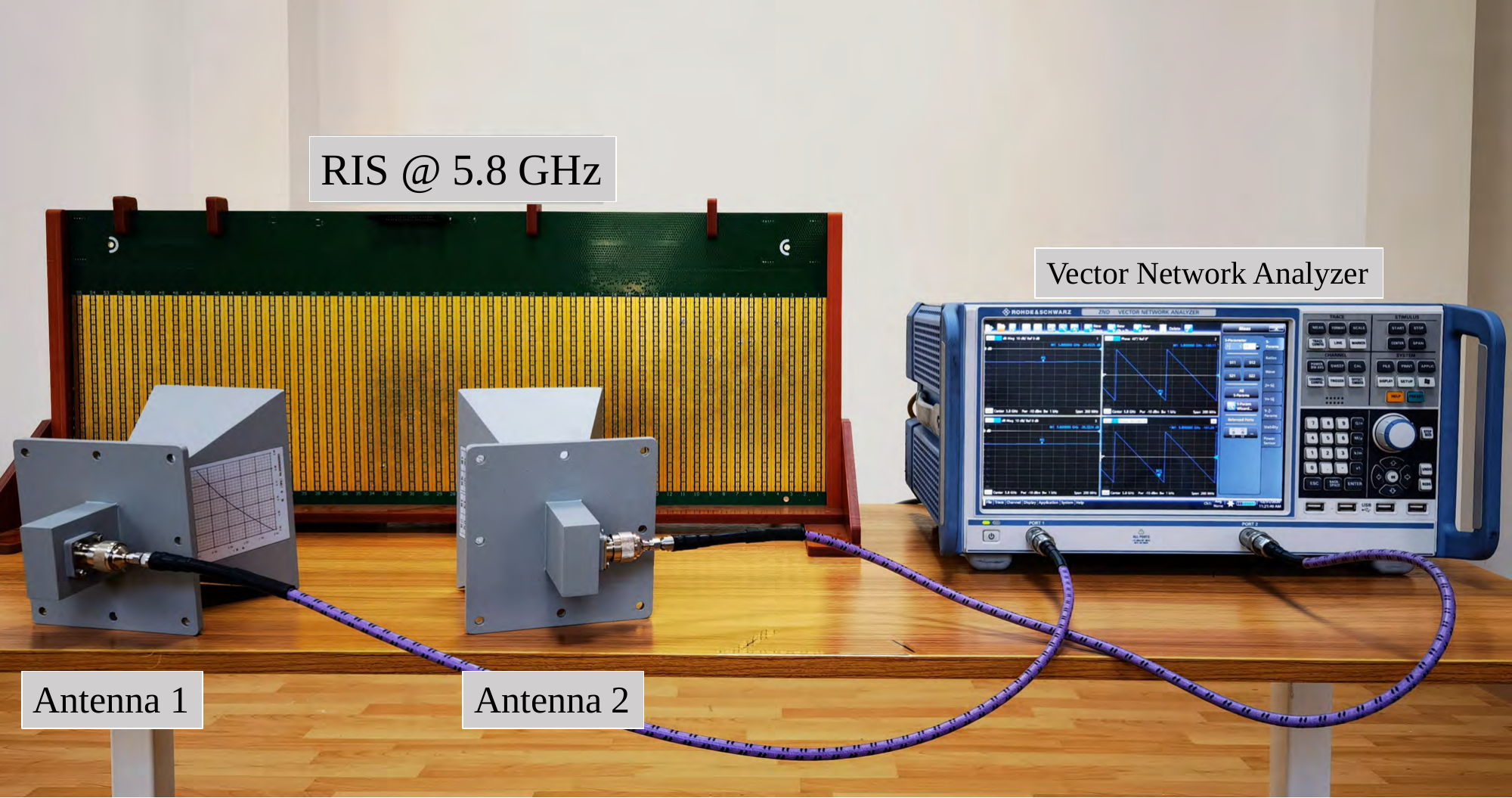}
	\caption{Channel reciprocity and reflection coefficient test environment.}
	\label{fig:gain_test}
\end{figure}

\begin{figure}[t!]
	\centering
	\subfloat[Amplitude vs. frequency]{\includegraphics[width=\linewidth]{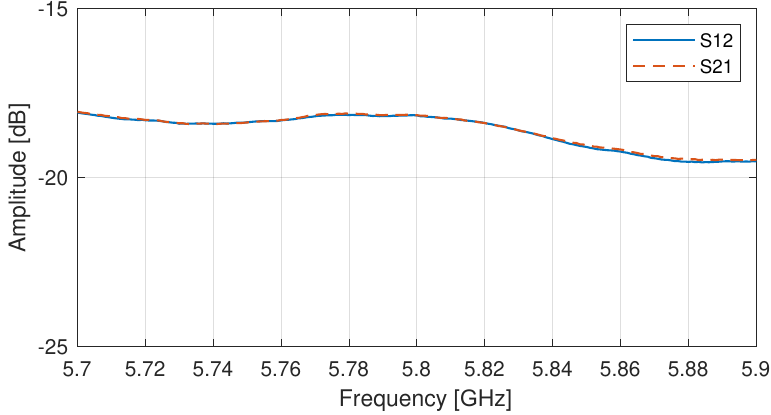}%
		\label{fig:channel_gain}}
	\hfil
	\subfloat[Phase vs. frequency]{\includegraphics[width=\linewidth]{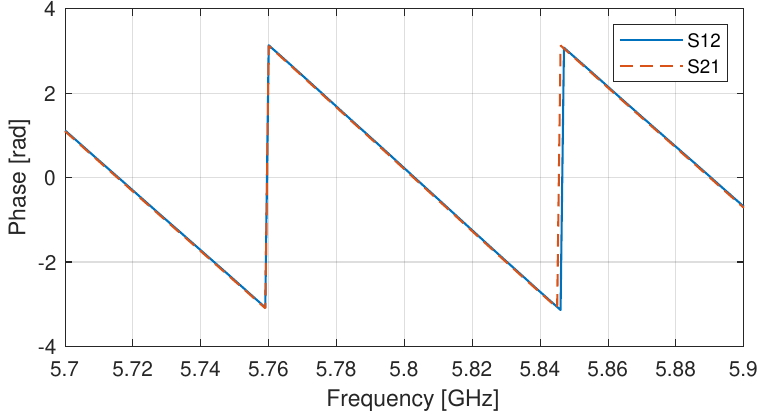}%
		\label{fig:channel_phase}}
	\caption{Measured S12 and S21 parameters of the RIS wireless channel, frequency range from 5.7 GHz to 5.9 GHz.}
	\label{fig:channel_reciprocity}
\end{figure}

When it comes to RIS-aided communications, UL/DL channel reciprocity would imply that the RIS can be configured to beamforming signals in UL and then work equally well in the DL without having to be reconfigured.
Motivated by this, we conduct experiments to verify the UL/DL channel reciprocity of RIS.
We use the vector network analyzer to test the S-parameters of the UL and DL. The test environment is shown in \figurename{~\ref{fig:gain_test}}, and two horn antennas are placed close in front of the RIS to minimize the effects of multipaths other than the one reflected by the RIS. 
A lot of other trials have been made with different configurations and antenna positions. However since they lead to the same conclusion, we only show one example of the measured S12 and S21 parameters of the RIS wireless channel in  \figurename{~\ref{fig:channel_reciprocity}}.
It can be seen from the figure that in the amplitude-frequency curve and phase-frequency curve, the S12 and S21 parameters are quite consistent with each other. Hence, despite the incident-angle-dependent behavior, the UL and DL channels in RIS-aided communication are reciprocal in our case. 
It means a channel estimate computed in one TX-RX direction can be reused when transmitting in the opposite direction, provided that the configuration remains unchanged. It also implies that if we configure the RIS for one direction, then we may keep the configuration when transmitting in the opposite direction. 

\subsection{Radiation Pattern Test}

\begin{figure}[t!]
	\centering
	\includegraphics[width=\linewidth]{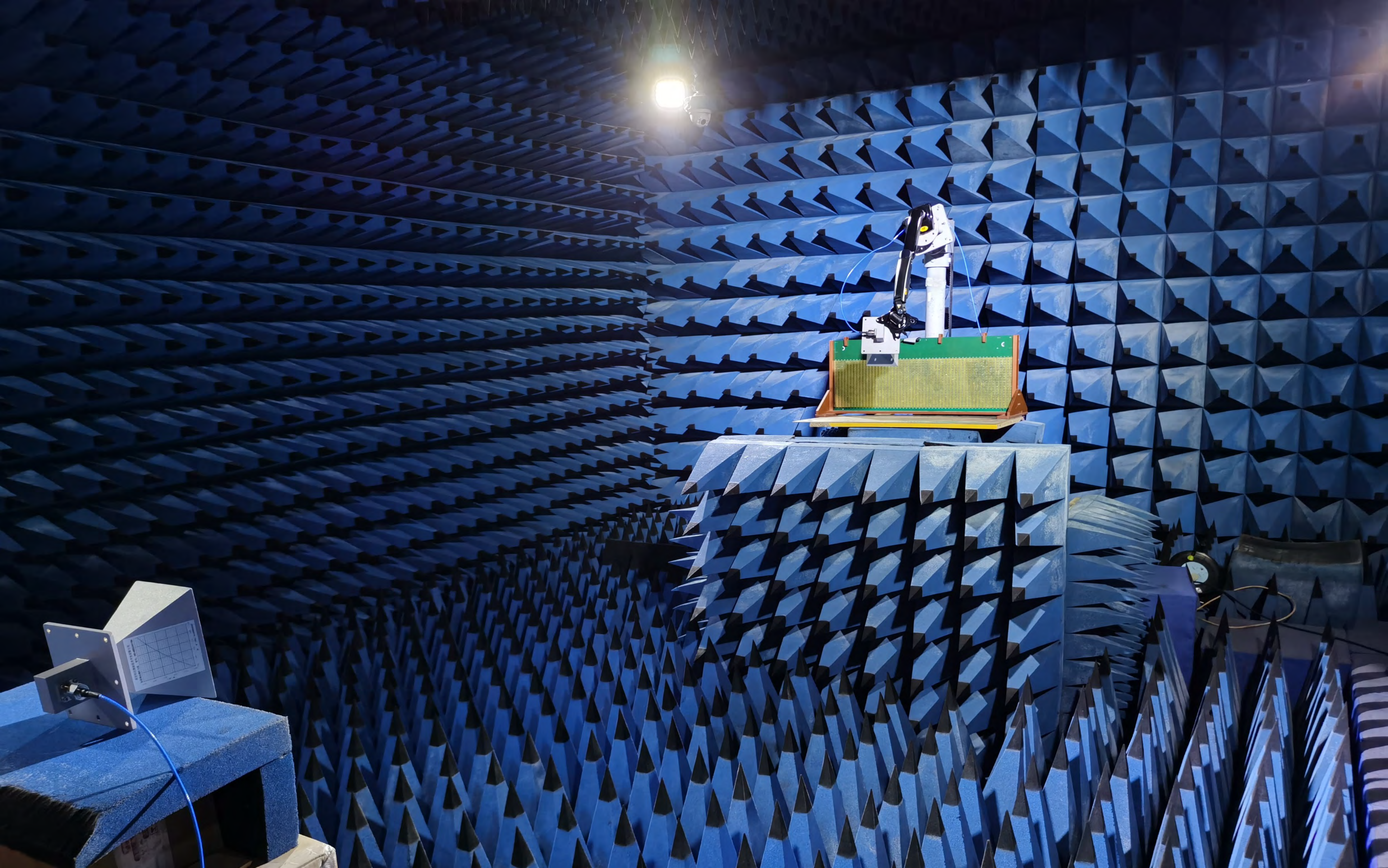}
	\caption{The microwave anechoic chamber of size $ 4 \, \mathrm{m} \times 6 \, \mathrm{m}\times 4 \, \mathrm{m}$. The RIS-RX antenna distance is 4.8 m. The TX-RIS distance is 40 cm.}
	\label{fig:mac}
\end{figure}

Next, we test the beamforming performance of the fabricated RIS.
The experiment is carried out in a microwave anechoic chamber as shown in \figurename~\ref{fig:mac}. 
The RIS and transmit antenna are fixed on a rotating platform, with the transmit antenna facing the RIS. 

\begin{figure}[t!]
	\centering
	\includegraphics[width=\linewidth]{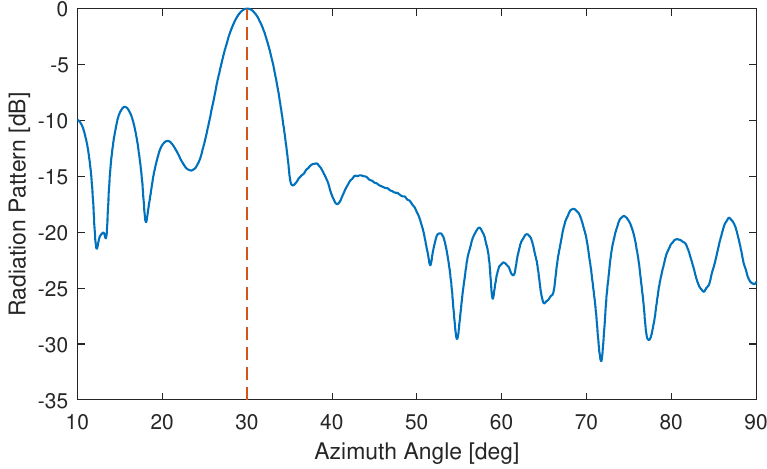}
	\caption{The measured radiation pattern of the RIS with pre-defined reflection coefficients.}
	\label{fig:30degree_greedy}
\end{figure}

We adjusted the reflection coefficients according to the 2D-DFT codebook described in Section~\ref{sec:codebook} to form a beam in the azimuth direction of 30 degrees. The codewords were 1-bit quantized, as previously described.
The platform was rotated to measure the radiation pattern. The result is shown in \figurename~\ref{fig:30degree_greedy}, where the gains are normalized so the maximum value is 0 dB.
The measurement shows that a high gain beam is generated in a direction of $ 30^\circ $ as expected.
The half-power beamwidth is $ 5.2 ^\circ $. The largest side lobes is $-8.79$ dB and is located to the left, while the largest sidelobe at the right side is $-11.8$ dB. 

\subsection{Indoor Over-the-air Test}

\begin{figure}[t!]
	\centering
	\includegraphics[width=\linewidth]{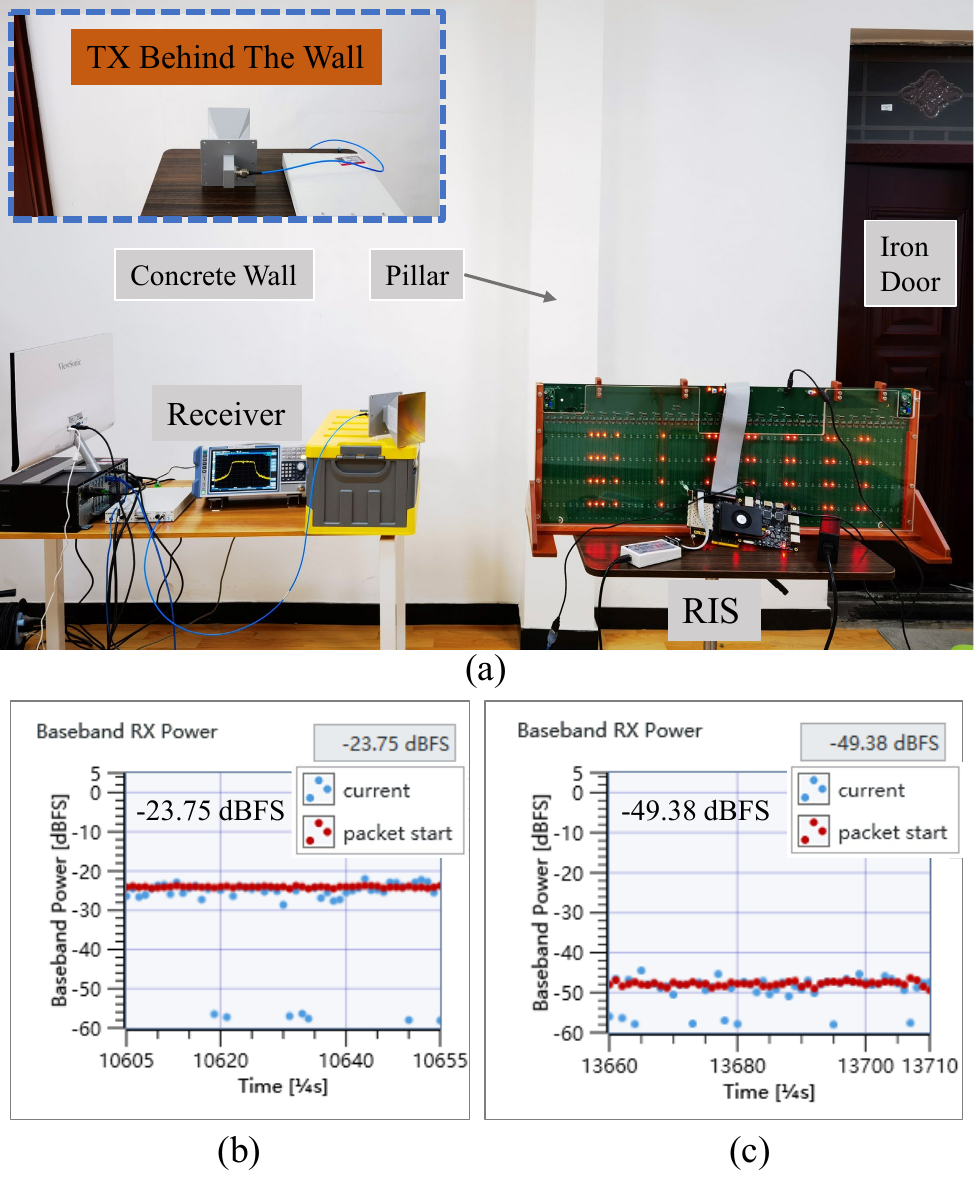}
	\caption{Indoor non-LoS test: (a) the testing scenario: the transmitter is behind a 30 cm concrete wall pointing at the RIS, the the thickness of reinforced concrete pillar is 53 cm, the iron door is closed, and the transmit power is -16 dBm; (b) the Baseband RX power is around $-23.75$ dBFS when using RIS; (c) the Baseband RX power is $-49.38$ dBFS when using a copper plate.}
	\label{fig:through_wall}
\end{figure}

We will now present results from the over-the-air  test that was made in the indoor environment shown in \figurename{~\ref{fig:through_wall}} (a).
The transmitter is placed in the corridor and the receiver is in a room next to it.
The transmitter and receiver are separated by a 30 cm thick concrete wall, which includes a 53 cm thick pillar in the middle.
In such a case without an LoS path, the transmitted signal suffers from penetration loss before reaching the RIS.
Nevertheless, the proposed Algorithm \ref{alg:GreedyAlgorithm} can be utilized to configure the RIS to gradually increase the received signal power. Note that the RX received power is utilized during the feedback. It is not perfect due to the influence of noise, interference, and time-varying channel. However, we may average the received power over time in order to make the algorithm more stable.
\figurename{~\ref{fig:through_wall} (b)} shows the received baseband power when using the RIS (after Algorithm \ref{alg:GreedyAlgorithm} has finished). \figurename{~\ref{fig:through_wall} (c)} shows the corresponding received power in the reference case where the RIS is replaced with a copper plate (shown in \figurename{~\ref{fig:50mtest} (b)}). Note that the position and angle of the copper plate are the same as the RIS. The curve marked with ``packet start" is the received power when a packet is successfully decoded, while the ``current" curve is the power at current moment, which may not have data transmission.
The results show that an RIS configured using the proposed algorithm brings a power gain of around 26 dB.
While it is generally recognized that the RIS technology is effective in LoS scenarios \cite{RIS_SPMAG}, our experiment shows that an RIS can be very effective also in non-LoS scenarios, at least over short distances.

\subsection{Outdoor Over-the-air Test}

We will now describe our outdoor field trials.
The testing sites were chosen at the campus of Huazhong University of Science and Technology (HUST).
Table \ref{tab:outdoorparameters} summarizes the main parameters of the system settings.
We chose two propagating distances for the field trial: 50 m and 500 m.

\begin{table}[t!]
	\renewcommand{\arraystretch}{1.3}
	\caption{System Parameters of Outdoor Test}
	\label{tab:outdoorparameters}
	\centering
	\begin{tabular}{|c|c|}
		\hline\hline
		\textbf{Parameter} &  \textbf{Value}  \\ \hline
		 Center frequency  &     5.8 GHz      \\ \hline
		Modulation method  &       OFDM       \\ \hline
		    Bandwidth      &      20 MHz      \\ \hline
		  TX-RX distance   &  50 m or 500 m   \\ \hline
		  Transmit power   & 13 dBm or 23 dBm \\ \hline\hline
	\end{tabular}
\end{table}

\subsubsection{50 meter test}

\begin{figure}[t!]
	\centering
	\subfloat[]{\includegraphics[width=\linewidth]{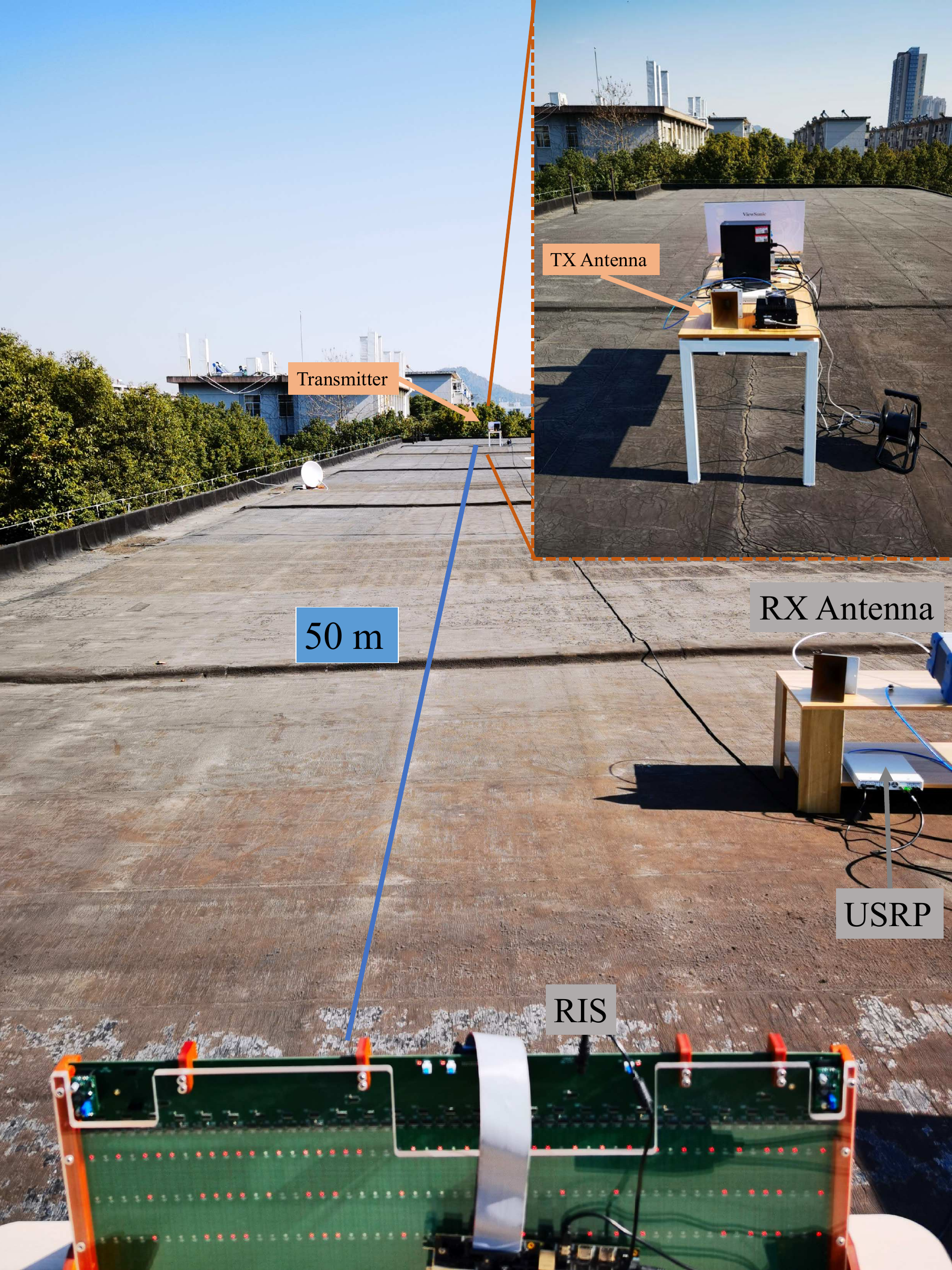}%
		\label{fig:50mphoto}}
	\hfil
	\subfloat[]{\includegraphics[width=\linewidth]{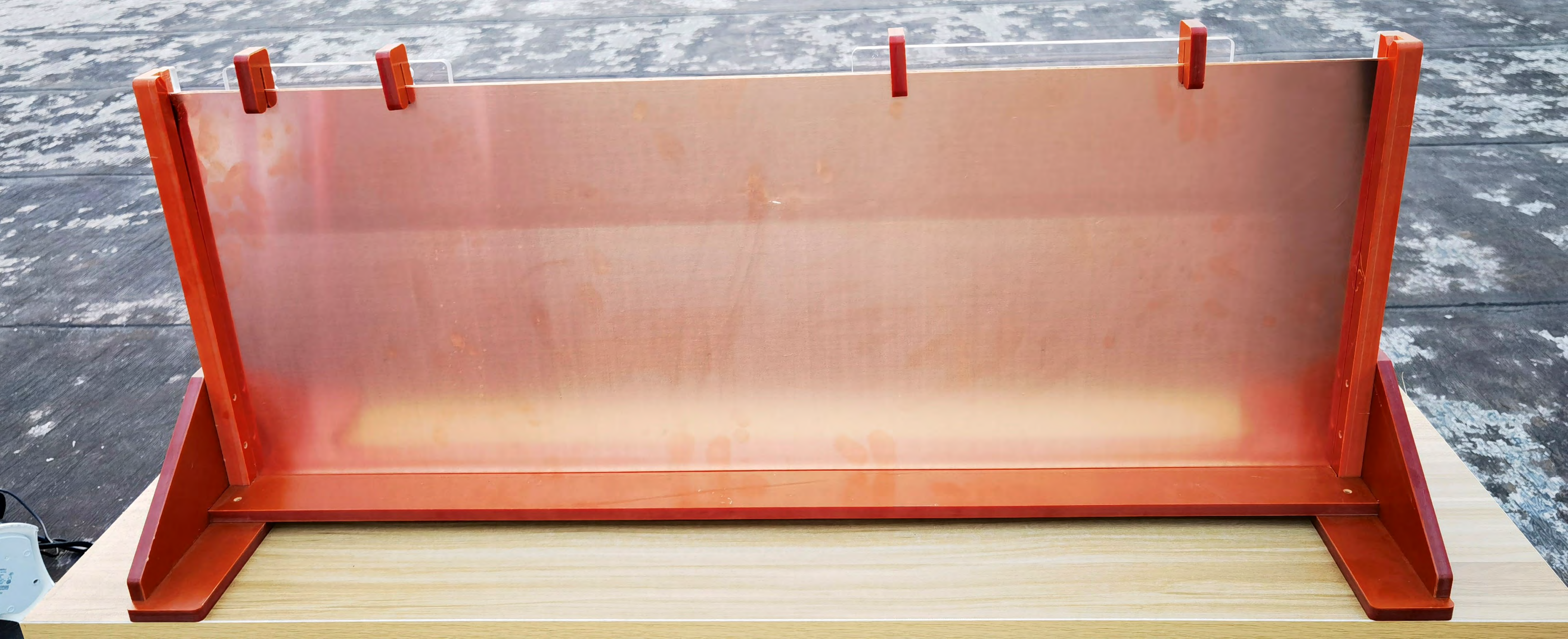}%
		\label{fig:copper}}
	\hfil
	\subfloat[]{\includegraphics[width=0.45\linewidth]{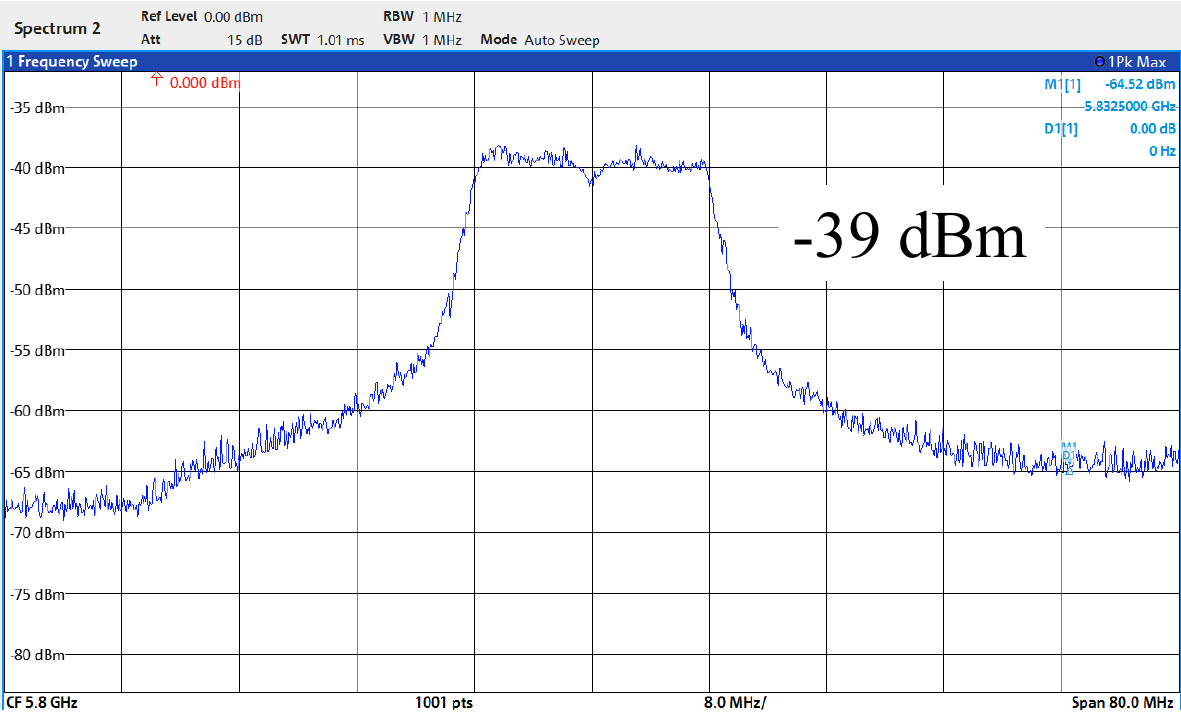}%
		\label{fig:50mgreedy}}
	\quad
	\subfloat[]{\includegraphics[width=0.45\linewidth]{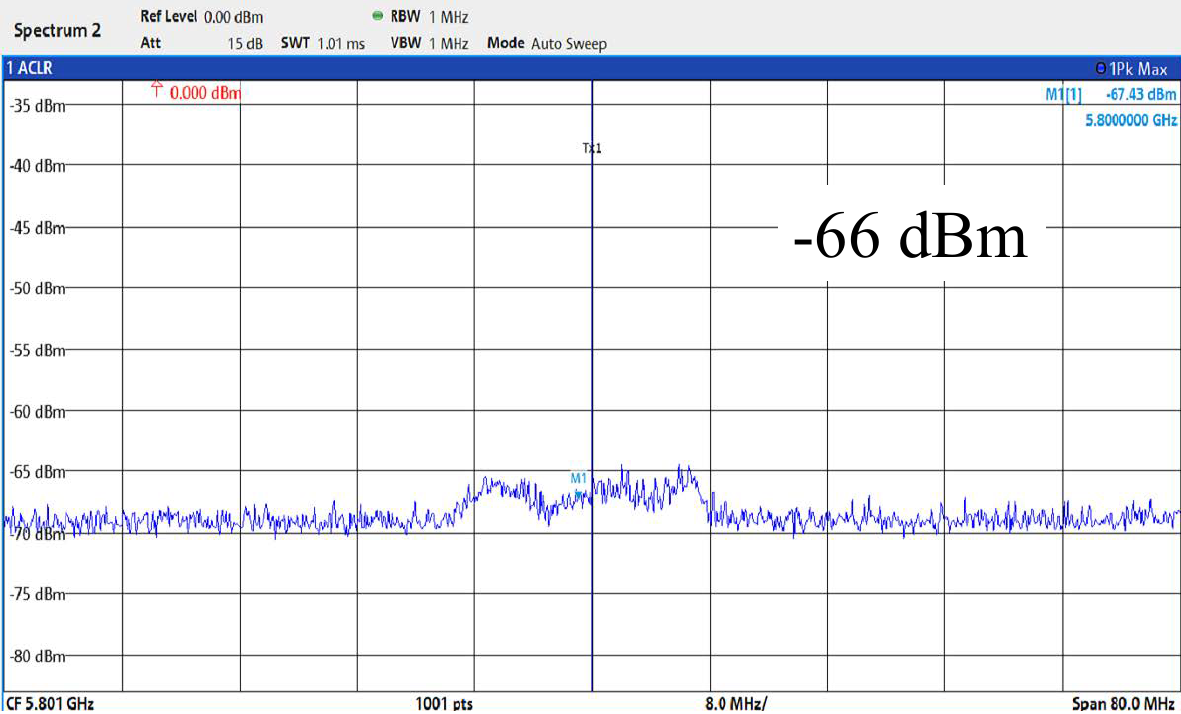}%
		\label{fig:50mcopper}}
	\caption{The 50 m outdoor over-the-air test: (a) the scene of the transmitter and receiver; (b) a copper plate of the same size as the RIS; (c) spectrum when using the RIS; (d) spectrum when using the copper plate.}
	\label{fig:50mtest}
\end{figure}

\figurename{~\ref{fig:50mtest} (a)} shows the environment, including the transmitter, receiver, and RIS locations, in the 50 m experiment. This field trial was carried out on the roof of our laboratory. The transmit power was 13 dBm. 
\figurename{~\ref{fig:50mtest} (c)} and \figurename{~\ref{fig:50mtest} (d)} show the spectrum of the receiver with and without RIS, respectively. The spectrum in
\figurename{~\ref{fig:50mtest} (c)} was measured at the receiver side using a spectrum analyzer in the case that the RIS had been configured using our proposed Algorithm \ref{alg:GreedyAlgorithm}.
\figurename{~\ref{fig:50mtest} (d)} shows the spectrum when the RIS is replaced with a copper plate as shown in \figurename{~\ref{fig:50mtest} (b)}.
{The results in \figurename{~\ref{fig:50mtest} (c)} and \figurename{~\ref{fig:50mtest} (d)} show a 27 dB power gain when using the RIS, as compared to using the copper plate.}
This number is well aligned with the power gain observed in the indoor test.

It is generally hard to compare measurement results like this with theory since there are many sources of uncertainty.
However, to put the numbers into perspective, suppose the copper plate would behave similarly to an RIS that has a random configuration. In that case, a fully optimized RIS beamforming could provide an average power gain up to $1100 = 30.4$ dB. Furthermore, it is known that 1-bit RIS configurations suffer a $-3.9$ dB loss on average \cite{Wu2020b}. Hence, the predicted power gain in this setup is 26.5 dB, which is very much aligned with the measurement results. While this is not an exact calculation, it gives a first-order indication that the RIS prototype with the proposed algorithm performs according to theory.

\subsubsection{500 meter test}

\begin{figure}[t!]
	\centering
	\includegraphics[width=\linewidth]{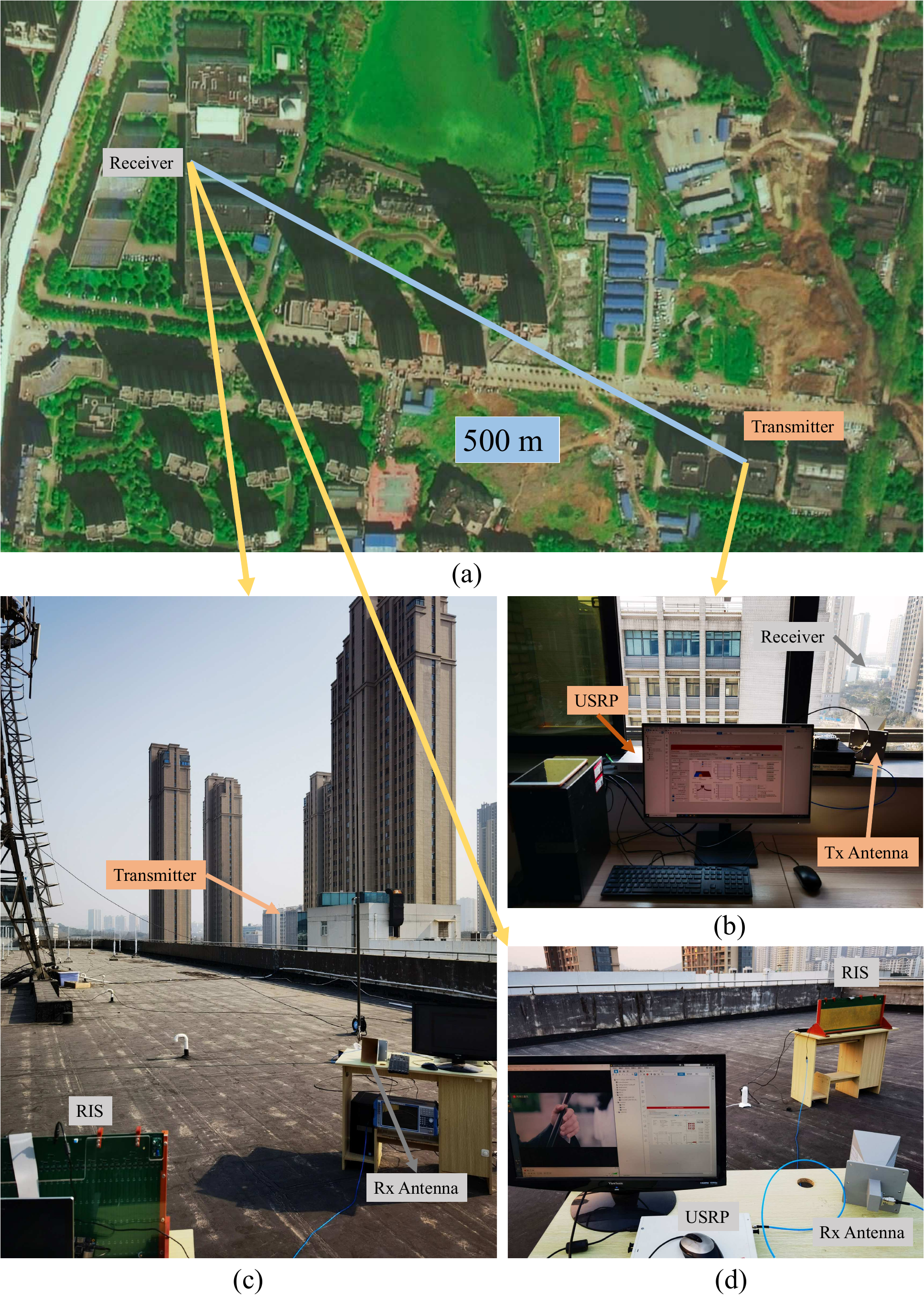}
	\caption{The 500 m outdoor over-the-air test: (a) location of the transmitter and receiver on the map; (b) the transmitter; (c) the site of the receiver; (d) a photo of the receiver when playing a live video.}
	\label{fig:500mtest}
\end{figure}

\begin{figure}[t!]
	\centering
	\includegraphics[width=\linewidth]{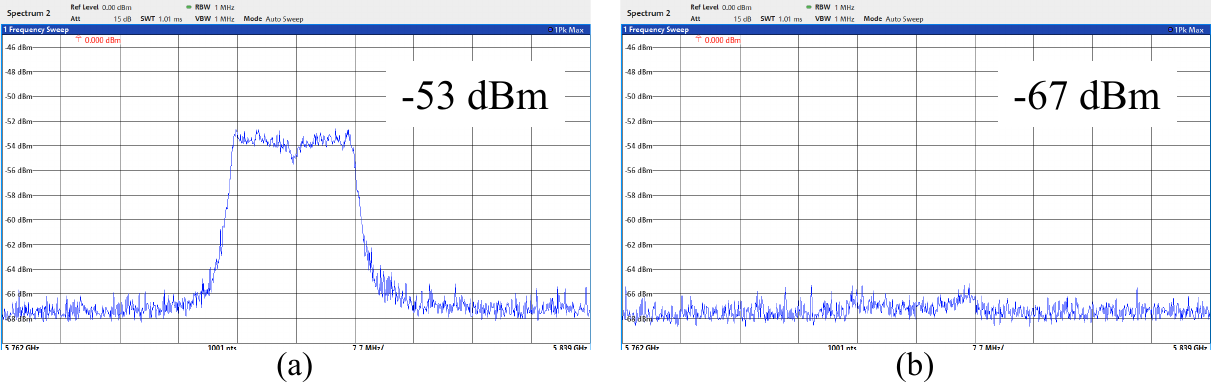}
	\caption{The spectrum of 500 m outdoor over-the-air test: (a) using the RIS; (b) using the copper plate.}
	\label{fig:500m_ppy}
\end{figure}

\figurename{~\ref{fig:500mtest}} shows the environment of our 500 m field trial, which we believe is the longest distance that has been considered so far in the literature on RIS-aided communications.
The experiment was carried out between two buildings. The transmit power was 23 dBm. 
Due to the long distance and limited transmit power, high-order modulation was not supported.
The maximum transmission rate in the measurement was 32.1 Mbps, which was achieved using 20 MHz of bandwidth and 16 QAM modulation.
\figurename{~\ref{fig:500m_ppy}} shows the spectrum of the received signal using the RIS (with the proposed algorithm) and using the copper plate. In this experiment, the power gain is 14 dB, which was sufficient to enable real-time  transmission of a video with $ 1920 \times 1080 $ resolution. The video was only playing smoothly when using the RIS. 

The measured power gain is smaller than in the short-distance experiments. There are several possible explanations for this result.
In general, the total received power originates both from the path via the RIS (or copper plate) and the combination of the multi-paths that are not involving the RIS.
One possible explanation is that the RIS provides the same improvements over the copper plate as before, but the impact on the total received power is smaller since the longer distance makes the path via the RIS proportionally weaker compared to the surrounding multi-paths.
Another possibility is that the more complicated propagation environment makes the channels via the RIS frequency-selective, which effectively reduces the beamforming gain since no RIS configuration fits for the entire band.

Note that the gain of the RIS depends on multiple parameters\cite{ntontin2021reconfigurable}, which includes the field pattern of the RIS elements, the distances between the RIS and the antennas, the angles of incidence/reflection, the number of the reflection elements, etc. A more detailed study is provided in \cite{Wangzipeng2021}.

\subsection{Power Consumption}

Finally, we provide a detailed description of the power consumption of the fabricated RIS. The power consumed by the individual electronic components on the RIS board is shown in Table \ref{tab:PowerConsumption}.

\begin{table}[t!]
	\renewcommand{\arraystretch}{1.3}
	\caption{Power Consumption Breakdown for RIS Board}
	\label{tab:PowerConsumption}
	\centering
	\begin{tabular}{|c|c|}
		\hline\hline
		          \textbf{Name}           &  \textbf{Power}  \\ \hline
		Bidirectional voltage translators &     0.0138 W     \\ \hline
		        Level regulators          &     0.918 W      \\ \hline
		         Varactor diode           &    0.00176 W     \\ \hline
		         \textbf{Total}           & \textbf{0.934 W} \\ \hline\hline
	\end{tabular}
\end{table}

It is interesting to note that the varactor diodes consume little power, although they are large in number. 
This is because reverse bias voltages are applied to them.
Most of the power is spent on the chips of the level regulators, to enable continuous bias voltage adjustment. This is useful for experiments. However, in a commercial product, it may not be necessary and low-power alternatives are available. 
Apart from the RIS board, the high-end FPGA controller consumes 1.5 W of power in our prototype. Nevertheless, the proposed beamforming algorithm could also be implemented on an ordinary microcontroller. This would bring down the power consumption of the controller to around 10 mW. In summary, we predict that an equal-sized RIS designed for minimum power consumption could consume far below 1 W.

\section{Conclusion}\label{sec:conclusion}

In this paper, we described our design and fabrication of an RIS with 1100 controllable elements working in the 5.8 GHz band. It was used to prototype an RIS-aided wireless communication system, where each element has two states (1-bit RIS). 
We proposed a self-adaptive RIS beamforming configuration algorithm that is based on real-time feedback with the help of a RIS-UE link. 
The geometrical channel properties and 1-bit structure were exploited to minimize the number of feedback iterations. 
The algorithm is particularly suited for setups where there are LoS paths from and to the RIS (e.g., where the RIS has been deployed to overcome shadowing), and can be utilized in any frequency band, from sub-6 GHz to sub-terahertz.
The proposed algorithm can be added on top of existing wireless protocols since only standard received power measurements are utilized when selecting the configuration. We first evaluated basic propagation properties in an anechoic chamber.  We noticed that the reflection coefficients of the RIS elements are varying with the incident angle, verified that channel reciprocity holds, and demonstrated the effectiveness of the reconfigurable beam-steering capability by measuring the radiation pattern.

We also evaluated the prototype in indoor and outdoor field trials.
In the indoor scenario, the RIS provided a  26 dB power gain at the receiver, even when the signal is blocked by a 30 cm thick concrete wall. 
In the short-range outdoor test, the RIS provided a 27 dB power gain. These gains are well aligned with theory and demonstrate the effectiveness of the proposed beamforming algorithm.
We also carried out the world's first long-range outdoor field trial, over 500 m. The RIS provided a 14 dB gain, which was sufficient to deliver 32 Mbps using only 23 dBm of transmit power.
It should be noted that the power consumption of our RIS prototype is only about 1 W. Our results demonstrated the great potential of RIS for future communication systems.

\section*{Acknowledgment}

Discussions with Linglong Dai and Shi Jin are gratefully acknowledged.

\bibliographystyle{IEEEtran}
\bibliography{references}

\end{document}